\documentclass{jfm}
\usepackage{graphicx}
\usepackage{epstopdf, epsfig}
\usepackage{newtxtext}
\usepackage{newtxmath}
\usepackage{xcolor}
\usepackage{soul} 
\usepackage{float}
\usepackage{lineno}
\usepackage[normalem]{ulem}

\graphicspath{{./figures/}}

\newcommand\Fro{\mbox{\textit{Fr}}}

\shorttitle{Disk wakes in nonlinear stratification}
\shortauthor{Gola et al.}

\title{Disk wakes in nonlinear stratification}

\author{Divyanshu Gola \aff{1}, S. Nidhan\aff{1}, J. L. Ortiz-Tarin\aff{1}, Hieu T. Pham\aff{1}, S. Sarkar\aff{1}\corresp{\email{sarkar@ucsd.edu}}}

\affiliation{\aff{1}Department of Mechanical and Aerospace Engineering, University of California San Diego, CA 92093, USA}	

\begin{document}
\maketitle
\begin{abstract}

Nonlinearity of density stratification  modulates buoyancy effects. We report results from a body-inclusive large eddy simulation (LES) of a wake in nonlinear stratification, specifically for a circular disk at diameter-based Reynolds number ($\Rey$) of $5000$. Five density profiles are considered; the benchmark has linear stratification and the other four have hyperbolic tangent profiles to model a pycnocline.
 The disk moves inside the central core of the pycnocline in two of those four cases and, in the other two cases with {a} shifted density profile, the disk moves partially/completely outside the pycnocline. The maximum buoyancy frequency ($N_{max}$) for all the profiles is the same. The first part of the study investigates the centered cases. Nonuniform stratification results in increasing wake turbulence relative to the benchmark owing  to reduced suppression of turbulence production as well as the wave trapping in the pycnocline. Steady lee waves are also  quantified to understand limitations of linear theory. The second part pays attention to the effect of a relative shift between the pycnocline and the disk. The wake defect velocity decays faster in the cases with a shift. The effect of disk location on the Kelvin wake waves (a family of steady waves within the pycnocline) and its modal form is obtained and explained by solving the Taylor-Goldstein equation. The family of unsteady internal gravity waves that are generated by the wake is also studied and the effect of disk shift is quantified.
 
\end{abstract}

\section{Introduction}








A pycnocline is a layer of fluid whose density changes rapidly with depth and so does the stratification described by the buoyancy frequency ($N$), where $N^2 = - (g/\rho_0) \partial \rho /\partial z$. Such layers in the ocean or atmosphere  affect  environmental turbulence and waves, the transport of tracers (nutrients, pollutants, etc.) and  the interaction of the environment with engineered structures.  The upper-ocean pycnocline is  flanked by a mixed layer on the top and a deeper  layer at the bottom. The mixed layer and the deep layer can also be stratified but their stratification levels are typically lower than that in the pycnocline. Since much of our  knowledge of turbulence in stratified wakes is derived from canonical wakes in  uniform stratification, it  {is} useful to study the wake of a canonical body inside and near a pycnocline where the change in stratification is rapid and nonlinear.

Most of the  past work on the flow features of bodies moving through a pycnocline has  involved  the internal wave structure. \cite{robey_thermocline_1997} studied internal waves generated by a sphere moving below a pycnocline  using experimental and numerical techniques for a wide range of body-based $\Rey$ and $\Fro$. \cite{nicolaou_internal_1995} experimentally studied the waves of an accelerating sphere in a thermocline.
The phase configuration for two-dimensional trapped internal waves was theoretically and experimentally studied by \cite{stevenson_phase_1986} for a cylinder moving in a thermocline generated using salt brine. Often, these experimental observations of wave patterns are compared to the theoretical results of earlier studies, e.g.,  \cite{barber_dispersion_1993} and \cite{keller_internal_1970},  {where} analytical results involving the dispersion relation of internal waves and their modal wave forms for nonlinear stratification profiles are provided.  

A hyperbolic tangent profile {bridging two regions} with different values of $N$ has been often used to model nonlinear density variation within and density jump  across a pycnocline. \cite{ermanyuk_force_2002, ermanyuk_force_2003} used a hyperbolic tangent profile to study forces on an oscillating cylinder and sphere, while \cite{grisouard_generation_2011} used a hyperbolic tangent profile with a constant bottom $N$ to study internal solitary waves in a pycnocline using direct numerical simulation. A hyperbolic profile of $N$ has been used for studying a weakly stratified shear layer adjacent to a uniformly stratified region \citep{pham_dynamics_2009} as well as an asymmetrically stratified jet \citep{pham_internal_2010}.

There have been numerous studies on evolution of turbulent wakes under stratification.
 An early study of  Froude number $O(10-1000)$ wakes by \cite{lin_wakes_1979} showed that stratification starts to affect the wake at buoyancy time scale ($Nt$) of $O(1)$ resulting in a non-axisymmetric evolution. The multistage wake decay was later characterized by \cite{spedding_evolution_1997} into three separate regimes, namely  three-dimensional (3-D), non-equilibrium (NEQ), and quasi-two-dimensional (Q2D) and quantitative turbulence measurements were obtained in this and  following laboratory studies.

Temporal simulations have enabled longer simulations in terms of evolved time ($T$)  {or streamwise distance ($x$) using the transformation} $T = x/U_{\infty}$, where $U_{\infty}$ is the free-stream velocity. \cite{gourlay_numerical_2001} found the appearance of pancake vortices in the Q2D regime in their temporal DNS of a wake at $\Rey = 10^4$ and $\Fro = 10$. \cite{brucker_comparative_2010} in their DNS of towed and self-propelled wakes at $\Rey = 10^4$ and $5 \times 10^4$ quantified the wake turbulence statistics and found that the higher defect velocity in stratified wakes is a result of the buoyancy induced reduction of the Reynolds shear stress and ultimately the turbulent production term. This results in a longer lifetime of the stratified wake, also verified later by  \cite{redford_numerical_2015} in their DNS of a weakly stratified turbulent wake. They also found that the horizontal nature of the wake in Q2D regime resulted in the dominance of  the lateral Reynolds shear stress so that the decay of the mean wake velocity was faster (compared to NEQ regime) and was accompanied by an increase in the turbulent kinetic energy. Other studies that use temporal wake models include \cite{dommermuth_numerical_2002}, \cite{diamessis_spectral_2005}, \cite{de_stadler_simulation_2012} and \cite{abdilghanie_internal_2013}.

Temporal simulations have been shown to be sensitive to the choice of initial conditions \citep{redford_universality_2012}. Besides, inclusion of vortex shedding at the body and  lee wave generation presents a challenge to temporal models. In recent work on stratified wakes, these limitations have been overcome by body-inclusive simulations. \cite{orr_numerical_2015} conducted a numerical study of a sphere wake at $\Rey = 200$ and $1000$ where they identified  vortex shedding and lee waves. \cite{pal_regeneration_2016} simulated the wake of a sphere at $\Rey = 3700$ using DNS and found that there is a resurgence of turbulent fluctuations below a critical $\Fro$ instead of a monotone suppression. This observation was further examined by  \cite{chongsiripinyo_vortex_2017} who found that, despite the low $\Fro$,  vortex stretching was dominant in the near wake,  resulting in  small-scale, three dimensional turbulence. The decay of a disk wake at $\Rey = 5\times10^4$ was examined  by \cite{chongsiripinyo_decay_2020} who  decomposed its evolution into three regimes based on buoyancy time scale and horizontal Froude number of the fluctuations (instead of the mean wake): weakly stratified, intermediately stratified and strongly stratified turbulence. Turbulence scales were also used to characterize wake transitions by \cite{zhou_large_2019}. \cite{ortiz_spheroid_2019} performed LES of flow past a prolate spheroid with laminar boundary layer separation into the near and intermediate wake and found that {at $\Fro \sim O(1)$,} buoyancy effects are stronger for slender bodies as compared to bluff bodies.

Interaction of the turbulent wake with the stratified background leads to the generation of unsteady internal waves. \cite{abdilghanie_internal_2013} found that the NEQ regime is prolonged at high $\Rey$, resulting in internal wave radiation that persists up to $Nt \approx 100$. \cite{brucker_comparative_2010} showed that internal wave flux dominates the turbulent dissipation during $20 < Nt < 75$ for a wake at $\Rey = 5\times10^4$ and $\Fro = 4$. In contrast, \cite{redford_numerical_2015} showed that for very high \Fro, the internal wave activity, although still more pronounced at the start of the NEQ regime, makes a negligible contribution to the turbulent energy budget. \cite{rowe_internal_2020} characterized the angles for the strongest internal waves and found that after $Nt = 10$, internal wave radiation is an important sink for wake kinetic energy.  \cite{nidhan_spectral_2022} used spectral proper orthogonal decomposition and connected the leading wake eigenmodes to the wave field. Besides numerical techniques, many experimental \citep{gilreath_experiments_1985, hopfinger_internal_1991, chomaz_structure_1993, bonneton_internal_1993, brandt_internal_2015, meunier_internal_2018}  as well as theoretical \citep{sturova_wave_1974, voisin_internal_1_1991, voisin_internal_2_1994, voisin_iwb_2003, voisin_sphere_2007} studies have been done to analyze the internal wave field in a stratified flow. 

Study of the wake in the vicinity of a pycnocline with a characterization of turbulence and its  interaction with trapped internal waves is limited. \cite{pham_internal_2010} studied the interaction between internal waves from a shear layer and an adjacent stratified jet, and classified waves that are trapped/reflected by the jet and waves that transmit through the jet. \cite{sutherland_internal_2004} and \cite{sutherland_internal_2016} calculated analytical expressions for internal wave transmission through density staircases for a stationary, two-dimensional Boussinesq fluid.  \cite{voropayeva_numerical_1997}  simulated a temporally evolving wake in  a nonlinear stratification using a Reynolds-averaged turbulence model.
 To the best of the authors' knowledge, the present study is the first body-inclusive wake LES  for a nonlinearly stratified fluid. We aim to assess the effect of nonlinear stratification on wake turbulence, its interaction with trapped waves as well as characterize the far-field lee waves. 
  The disk is chosen as a canonical generator of a bluff-body wake  to avoid the computational expense incurred in resolving the boundary layer that develops on a long body.



Section \ref{methodology} {describes} the numerical setup and the simulation parameters. The results are divided into two sections. The basic differences between linear and nonlinear stratification when the disk is centered in the pycnocline layer are studied in section \ref{part1}. The effect of  relative shift between the density profile and the disk, so that the disk is partially/completely outside the pycnocline layer, is discussed in section \ref{part2}.  {We summarise and conclude the study in section} \ref{conclusions}.

\section{Methodology}\label{methodology}

\subsection{Governing equations and numerical scheme}\label{numerical_details}

\begin{figure}

\centerline{\includegraphics[width=0.8\linewidth, keepaspectratio]{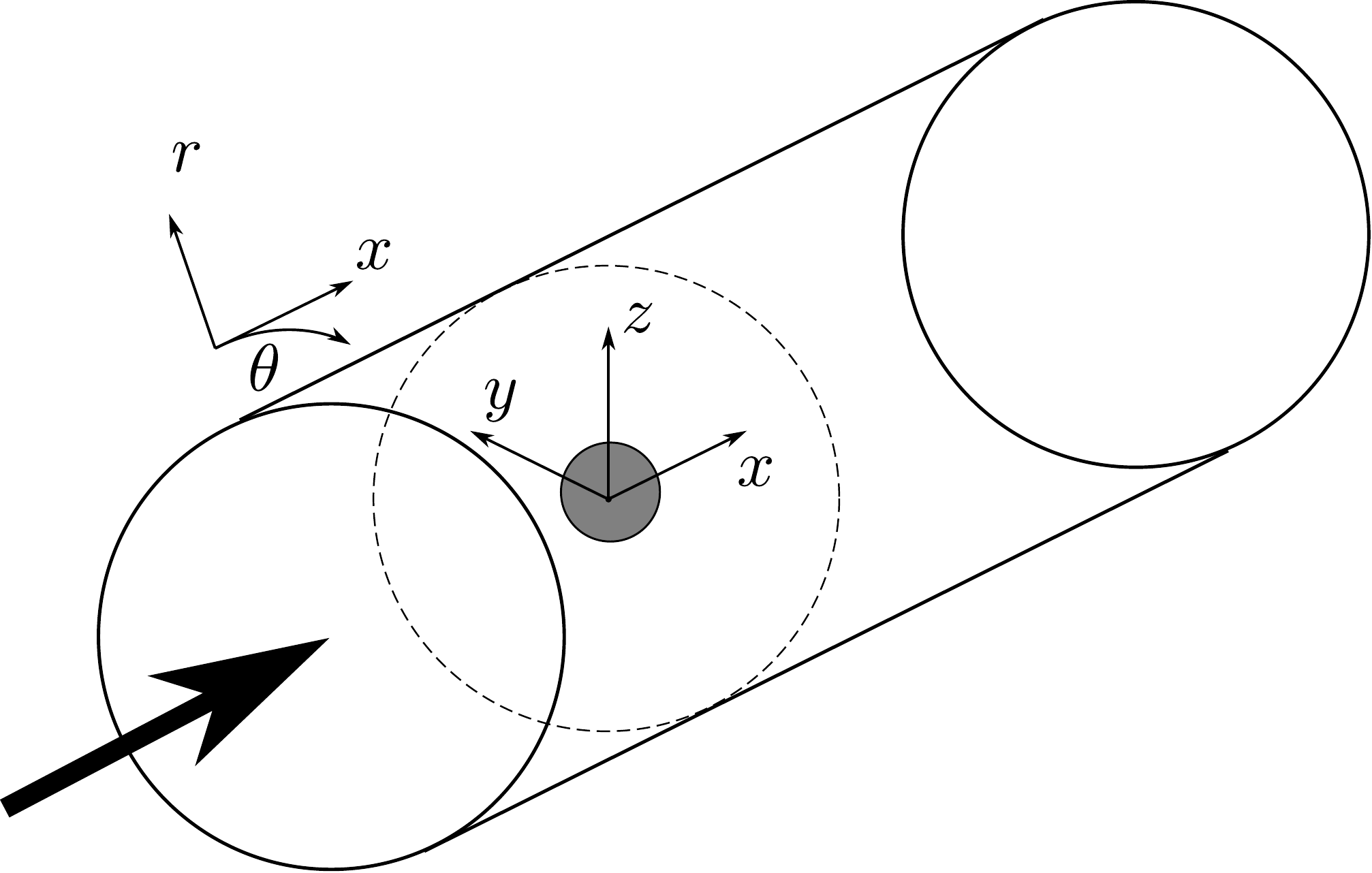}}
\caption{Schematic showing the simulation setup and domain.}
\label{fig:simsetup}
\end{figure}

The wake of a disk is simulated by solving the three dimensional, incompressible, unsteady form of the conservation equations for mass, momentum and density. A high-resolution large eddy simulation (LES)  with the Boussinesq approximation for density effects is used. The disk is immersed perpendicular to a flow with velocity $U_{\infty}$. The equations are numerically solved in cylindrical coordinates but both Cartesian $(x,y,z)$ and cylindrical $(r,\theta,x)$ coordinates are appropriately used in the discussion. Here, $x$ is streamwise, $y$ is spanwise and positive along $\theta=0^{\circ}$, and $z$ is vertical and positive along $\theta=90^{\circ}$ (Fig. \ref{fig:simsetup}). The density field $(\rho(\mathbf{x},t))$ is split into a constant reference density $(\rho_{0})$, the variation of the background $(\Delta \rho_{b}(z))$, and the flow induced deviation, $(\rho_{d}(\mathbf{x},t))$ so that $\rho(\mathbf{x},t)=\rho_{0}+\Delta \rho_{b}(z)+\rho_{d}(\mathbf{x},t)$. The Reynolds number of the flow, defined as $\Rey = U_{\infty} D/\nu $ with $D$ being the diameter of the disk, is $5000$.   Since the stratification is non-uniform, the Froude number defined by $Fr (z) = U_{\infty}/N(z) D$ is variable and its case-dependent behavior is described in section \ref{profiles}. The minimum value of the Froude number ($Fr_{min}$)  is set to $1$ for all the cases. 

The filtered nondimensional equations are as follows 

\begin{equation} 
\frac{\partial u_{i}}{\partial x_{i}} = 0,
\label{conservation_eqn}
\end{equation}

\begin{equation} 
\frac{\partial u_{i}}{\partial t} + u_{j}\frac{\partial u_{i}}{\partial x_{j}} = -\frac{\partial p}{\partial x_{i}} + \frac{1}{\Rey} \frac{\partial}{\partial x_{j}}\Big[(1 + \frac{\nu_{sgs}}{\nu}) \frac{\partial u_{i}}{\partial x_{j}}\Big] - \frac{\rho_{d}}{Fr_{min}^{2}} \delta_{i3}, 
\label{momentum_eqn}
\end{equation}

\begin{equation} 
\frac{\partial \rho}{\partial t} + u_{j}\frac{\partial \rho}{\partial x_{j}} = \frac{1}{\Rey Pr} \frac{\partial}{\partial x_{j}}\Big[(1 + \frac{\kappa_{sgs}}{\kappa})\frac{\partial \rho}{\partial x_{j}}\Big], 
\label{density_eqn}
\end{equation}
where $u_{i}$ refers to the filtered velocities in $x$, $y$, and $z$ directions for $i = 1, 2$, and $3$ respectively. $\nu_{sgs}$ and $\nu$ in Eq. (\ref{momentum_eqn}) are the subgrid-scale kinematic viscosity and the kinematic viscosity, respectively, while $\kappa_{sgs}$ and $\kappa$ in Eq. (\ref{density_eqn}) are the subgrid scale diffusion coefficient and the diffusion coefficient, respectively. The Prandtl number defined by $Pr=\nu / \kappa$ is set to one. No major qualitative influence of $Pr$ was found in the wake study by \cite{de_stadler_prandtl_2010} who varied $Pr$ between 0.2 and 7. 
The dynamic eddy viscosity model~\citep{germano_dynamic_1991} is employed to obtain $\nu_{sgs}$, following the implementation of \cite{chongsiripinyo_decay_2020},  and the subgrid Prandtl number is {also} set to unity to obtain $\kappa_{sgs}$.
The parameters used for nondimensionalization are as follows: free-stream velocity ($U_{\infty}$) for velocity, disk diameter ($D$) for length, advection time ($D/U_{\infty}$) for time, dynamic pressure ($\rho U_{\infty}^{2}$) for pressure, and characteristic change in background density  {($-D (\partial \Delta \rho_{b} / \partial z) |_{max}$)} for the flow induced density deviation. 

The periodicity in the azimuthal direction is leveraged in solving the discretised pressure Poisson equation in the predictor step, reducing it to a pentadiagonal system of linear equations, which is then solved using a direct solver \citep{rossi_parallel_1999}. The disk is represented by the immersed body method of \cite{balaras_modeling_2004} and \cite{yang_embedded-boundary_2006}. 
At the inlet boundary, a uniform stream of velocity ($U_{\infty}$) is imposed while an Orlanski-type convective boundary condition is used for the outflow \citep{orlanski_simple_1976}. Neumann boundary condition is imposed at the radial boundary of the domain for density as well as the three velocity components. To prevent spurious reflection of waves back into the domain, sponge layers are used at the inlet (streamwise length $10D$), outlet (streamwise length $10D$) and the cylindrical walls (radial length $15D$) of the domain to gradually relax the velocities and the density to their respective {background} values at the boundaries.

\subsection{Density profiles}\label{profiles}
\begin{figure}
\centering
\centerline{\includegraphics[width=\linewidth, keepaspectratio]{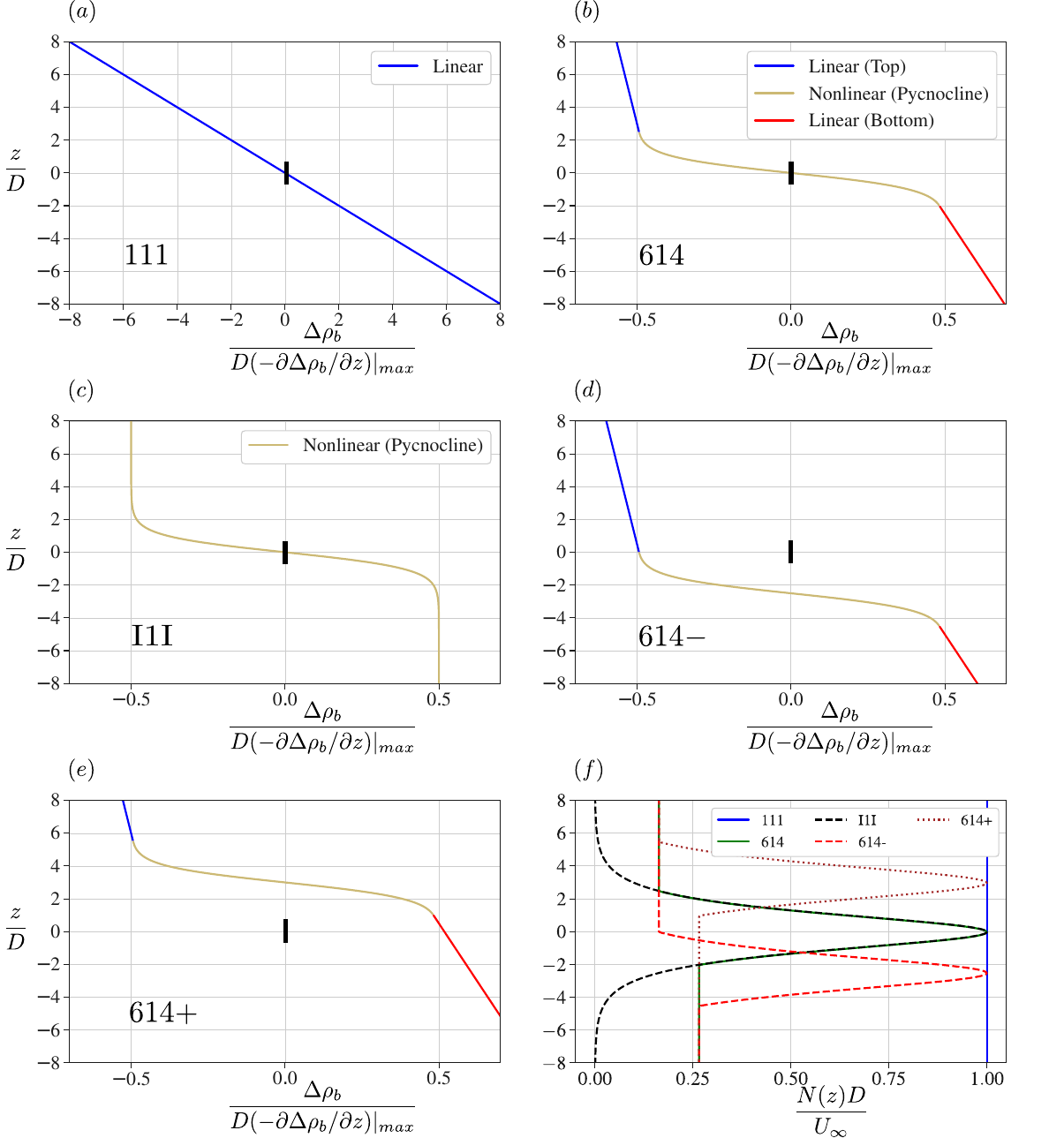}}
\caption{Variation of background density for (a) 111, (b) 614, (c) I1I, (d) $614-$, (e) $614+$, along with (f) respective buoyancy frequency profiles. Solid black line at the center of (a-e) represents the disk (not to scale). Profiles are summarized in table~\ref{tab:phys_parameters} and also discussed in text. For $614-$, the upper half of the disk is in a constant-$N$ region and the lower is in the pycnocline. For  $614+$, the  disk is in the bottom constant-$N$ region with its upper edge $0.5D$ below the pycnocline.
}
 
\label{fig:profiles}
\end{figure}

The profiles chosen for the variation of background density in the five cases are shown in Fig. \ref{fig:profiles}.    
 Except for the benchmark 111 case, the profiles have nonuniform $N(z)$ and $Fr(z) = U/N(z)D$.  Profile 111 with linear stratification has a constant  $Fr (z)$ which is the conventional body-based Froude number of   $\Fro = U_{\infty}/ND = 1$. The maximum value of the density gradient is the same for all four nonlinear profiles and corresponds to $Fr_{min} =1$. In Profile 614 (Fig. \ref{fig:profiles}b), a hyperbolic tangent function is used to bridge two linearly stratified regions (6 and 4 represent the value, rounded  to an integer, of the  local  $Fr$ in the linearly stratified regions).  Profile-I1I (I standing for ``Infinity") is a  complete hyperbolic tangent profile between two constant-density regions, therefore having infinite Froude number far above and below the body. 
 
 For the I1I and 614 profiles,  the disk center coincides with the center of the density profile and, furthermore,  the disk lies entirely within the nonlinearly stratified region. Profiles $614-$ and $614+$ are obtained after shifting profile 614 (and not the disk) vertically by $-2.5D$ and $3D$, respectively. The disk center is located at $z = 0$ for all five profiles and the negative and positive signs in the names $614-$ and $614+$ indicate whether the profile is shifted downwards or upwards, respectively.  The vertical  shift in $614-$ (Fig. \ref{fig:profiles}d) is chosen so that the upper half of the disk lies in the linearly stratified region with $\Fro = 6$ and the lower half in the pycnocline.
The  vertical shift in $614+$ (Fig. \ref{fig:profiles}e) is  chosen so that the disk lies entirely in  the linearly stratified region with $\Fro = 4$  while being very close to the pycnocline, specifically, its upper edge is $0.5D$  below the pycnocline. 
 
The  profile I1I is given by the following function, which was also used by \cite{ermanyuk_force_2002, ermanyuk_force_2003} and \cite{nicolaou_internal_1995} in their studies: 

\begin{equation} 
\Delta \rho_{b}(z)=-\frac{\rho_{0} \varepsilon}{2} \tanh{\frac{2z}{\delta}} .
\label{tanh_profile_function}
\end{equation}

The central nonlinear region of the 614 profile is obtained from the I1I profile.  The linearly stratified regions of  profile 614 are added by matching the slopes at $z=1.25\delta$ and $z=-\delta$ so that the density profile is continuously differentiable in $z$. This results in Froude numbers of $6.13$ and $3.74$ above and below the pycnocline layer respectively, for profile 614.

The  profiles $614-$ and $614+$ are shifted with respect to 614 and their central nonlinear region  is as follows:

\begin{equation} 
\Delta \rho_{b}(z)=-\frac{\rho_{0} \varepsilon}{2} \tanh{\frac{2(z - s)}{\delta}} ,
\label{tanh_profile_function2}
\end{equation}
where $s$ denotes the shift and takes values $-2.5D$ and $3D$ for $614-$ and $614+$, respectively.

The local Froude number corresponding to this pycnocline region can be calculated  using the local background buoyancy frequency: 
\begin{equation} 
Fr(z) = \frac{U_{\infty}}{\sqrt{\varepsilon gD}} \sqrt{\frac{\delta}{D}} \; \Big|\cosh{\frac{2(z-s)}{\delta}}\Big| .
\label{froude_number}
\end{equation}
The minimum value of $Fr(z)$ occurs at $z=s$. Thus, the minimum Froude number for profiles 614 and I1I is
\begin{equation} 
Fr_{min} = Fr(z=s=0) = \frac{U_{\infty}}{\sqrt{\varepsilon gD}} \sqrt{\frac{\delta}{D}} .
\label{fr_min}
\end{equation}

Eq. \ref{fr_min} highlights two important non-dimensional parameters for a body moving through a pycnocline:  $U_{\infty}/\sqrt{\varepsilon gD}$, which  is the conventional Froude number defined using reduced gravity $\varepsilon g$, and $\delta / D$, which is the nondimensional thickness of the pycnocline. In the simulations,  $U_{\infty}/\sqrt{\varepsilon gD} = 1/\sqrt{2}$ and  $\delta / D = 2$ for profiles 614 and I1I to obtain $Fr_{min} = 1$. Note that for 111, which has constant linear stratification throughout, $Fr_{min}=1$ still holds. {Parameters for each case 
are given in} Table~\ref{tab:phys_parameters}. 


\subsection{Domain and Grid}

The domain extends from $x/D = -L_{x}^{-} = -30$ to $x/D =  L_{x}^{+} =102$ in the streamwise direction and from $r/D = 0$ to $r/D =  L_{r} =60$ in the radial direction. The number of grid points employed for discretizing the domain is $N_{x}=2176$, $N_{\theta}=128$, and $N_{r}=479$ in the streamwise, azimuthal and radial directions respectively, resulting in approximately $130$ million grid points. The LES grid is non uniform in the streamwise and radial direction and is designed to have high resolution.  In terms of the Kolmogorov length ($\eta =(\nu^{3} / \epsilon_{k}^{T})^{1/4} $),  the maximum value of $\Delta x / \eta $ is $4.39$  and that of $\Delta r / \eta$ is $5.03$. Also, the turbulent dissipation rate used for the calculation of $\eta$ includes the resolved scale dissipation ($\epsilon = 2\nu \langle s_{ij}' s_{ij}' \rangle$) as well as the subgrid scale dissipation ($\epsilon_{sgs} = - \langle \tau_{sgs_{ij}}' s_{ij}' \rangle $ with $\tau_{sgs_{ij}}= -2 \nu_{sgs} S_{ij} $), where $s_{ij}'$ and $S_{ij}$ are the fluctuating and mean strain rates respectively. Most of the turbulent dissipation resides in the resolved scales.

\begin{table}

\begin{center}
\begin{tabular}{lcccccc }


Case  & $\frac{\delta}{D}$ & $\Fro_{min}$ & $\Fro_{z \rightarrow +\infty}$ & $\Fro_{z \rightarrow - \infty}$ & $\frac{s}{D}$ & Comment \\
\hline
$111$ & N/A            &  $1$  & $1$   & $1$           & $0$   & $N(z) = $ constant (Fig. 1a)  \\
I$1$I & $2$            &  $1$  & $\infty$   & $\infty$           & $0$   & $N(z \rightarrow \pm \infty) = 0$ (Fig. 1c)\\
$614$ & $2$            &  $1$  & $6.13$   & $3.74$           & $0$   & $N(z \rightarrow +\infty) = 0.163 $, $N(z \rightarrow -\infty) = 0.267$ (Fig. 1b)    \\
$614-$ & $2$            &  $1$  & $6.13$   & $3.74$           & $-2.5$   & $N(z)$ of $614$ shifted down  (Fig. 1d)  \\ 
$614+$ & $2$            &  $1$  & $6.13$   & $3.74$           & $3$   & $N(z)$ of $614$ shifted up  (Fig. 1e) \\


\end{tabular}

\end{center}

\caption{Physical parameters of the simulated cases. For each case, $\Rey = 5000$ and $Pr = 1$. The domain with $L_{r} = 60$, $L_{\theta} = 2\pi$, $L_{x}^{-} = 30$, and $L_{x+} = 102$ is discretised using $N_{r} = 479$, $N_{\theta} = 128$, and $N_{x} = 2176$ points.}
\label{tab:phys_parameters}
\end{table}

\section{Linear vs. nonlinear stratification}\label{part1}

\subsection{Steady lee waves}\label{lee_waves}

Disturbances in stratified environments generate internal waves. These can be steady lee waves generated by the body or unsteady internal waves generated by turbulence in the wake of body. In this section, the characteristics of the body generated lee waves on the vertical center plane for 111 and 614 will be analyzed using linear asymptotic theory. Note that since I1I is essentially unstratified away from the pycnocline layer, it does not show any lee waves.

\begin{figure}

\centerline{\includegraphics[width=\linewidth, keepaspectratio]{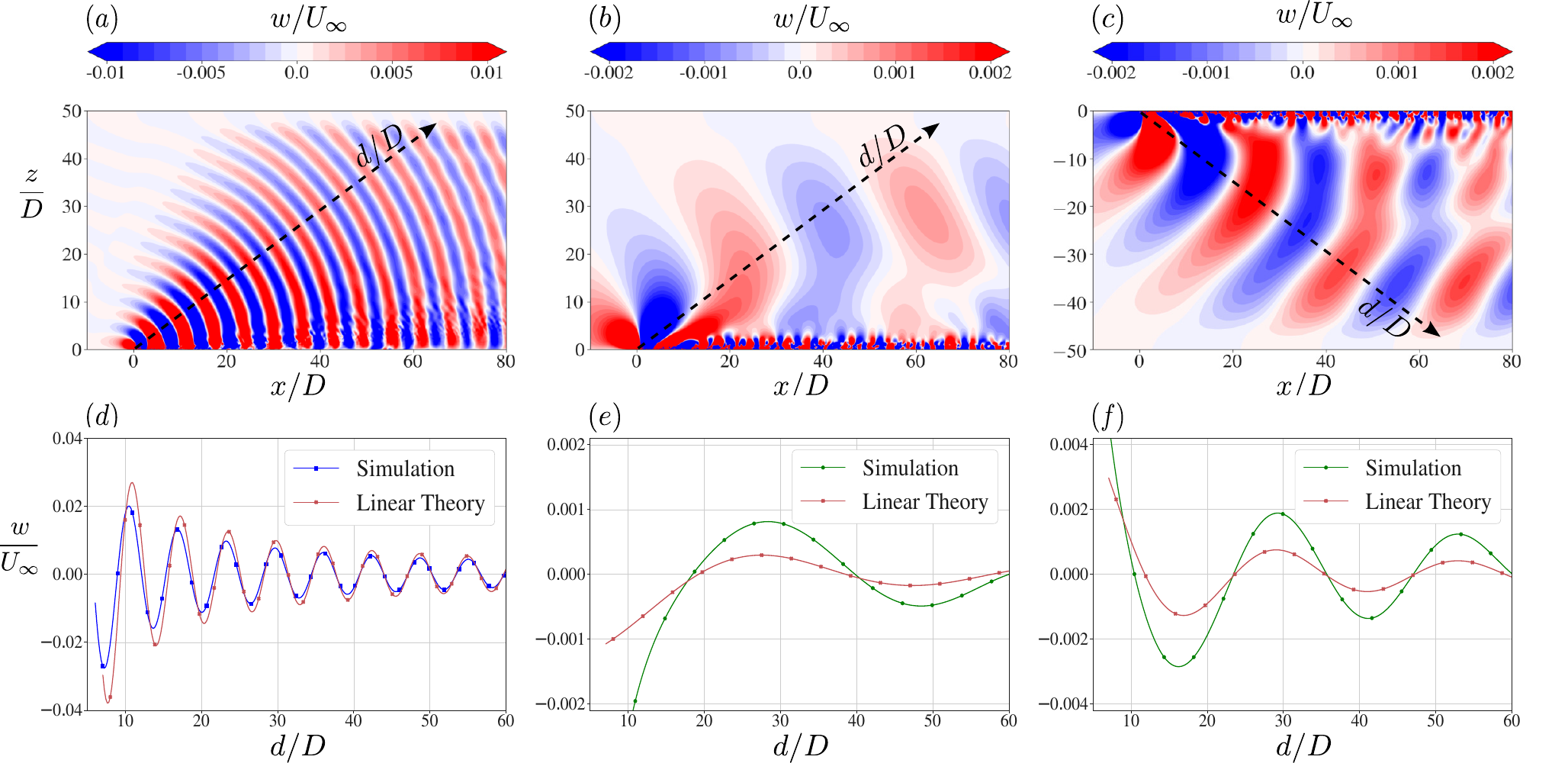}}
\caption{Vertical velocity contours showing lee waves and their amplitudes along the dashed arrow labeled by $d/D$. (a,d) - 111 ($90^{\circ}$ plane); (b,e) - 614 ($90^{\circ}$ plane); (c,f) - 614 ($270^{\circ}$ plane).
}
\label{fig:lee_waves}
\end{figure}

Instantaneous contours of vertical velocity ($w$) for 111 (top half) and 614 (top and bottom halves) on a vertical plane passing through the centerline are plotted in Fig. \ref{fig:lee_waves}(a-c). The {amplitude} of steady lee waves decays  moving away from the body. The lee waves observed in 111 are symmetric with respect to $z =0$  while  614 has two different sets of lee waves (top and bottom) owing to the two different local values of $Fr(z)$ {($6.13$ above and $3.74$ below)}  in the linearly stratified regions above and below the pycnocline layer. The waves in the  614 case have a much smaller amplitude than in the 111 case and a larger wavelength.

Linear theory, given by \cite{voisin_internal_1_1991, voisin_iwb_2003, voisin_sphere_2007} is used to analytically predict the wavelength and amplitude of the lee waves. The theory computes a wave function, $\chi$  from a linearized set of inviscid equations involving a source term $q$ on the right hand side of the continuity equation to model the moving body using a Green's function approach for large times ($Nt>>1$), which in our case translates to $x/D >> Fr$. Once $\chi$ is known, 
$w$ is  calculated. The expression for $w$ for the case of a Rankine ovoid as derived by \cite{ortiz_spheroid_2019} and employed for a 4:1 spheroid wake  can be reduced to give the wave pattern on the vertical centerline plane as:
\begin{equation}
w(x,y=0,z) \sim -\frac{mN}{\pi U_{\infty} r_{xz}} \cos{\psi} \sin{\Big(\frac{Na}{U_{\infty}} \cos{\psi}\Big)} \sin{\Big(\frac{N}{U_{\infty}} r_{xz}\Big)} ,
\label{w_lee_waves}
\end{equation}
 where $r_{xz}=\sqrt{x^{2}+z^{2}}$ and $\psi = \arctan(z/x)$. Also, $m$ and $a$ are calculated from the potential flow stagnation point solution for a Rankine ovoid of length $L_{ro}$ and cross-sectional diameter $D_{ro}$:
\begin{equation}
(L_{ro}^2 - 4a^2 )^2 = \Big(  \frac{8ma}{\pi U_{\infty}}  \Big) L_{ro}, \qquad D_{ro}^2 = \Big(  \frac{8ma}{\pi U_{\infty}}  \Big) \frac{1}{\sqrt{4a^2 + D_{ro}^2}}
\label{lro_dro}
\end{equation}
Note that \cite{ortiz_spheroid_2019} calculated $m$ and $a$ using the potential flow solution of a Rankine oval, which is a 2-dimensional solution. Eq. (\ref{lro_dro}) is based on the solution for a Rankine ovoid (which is 3-dimensional), and serves as a correction to \cite{ortiz_spheroid_2019}.
Note that, as  per Eq. \ref{w_lee_waves}, the wave amplitude decreases with decreasing $N$ or increasing $\Fro$.

The justification for using the expression for a Rankine ovoid instead of a disk is as follows. To apply linear theory to a bluff body such as a disk,  the separation bubble created by the disk  {should also be included} as a part of the extended body involved in wave generation. The complete extended body for 111, I1I and 614  cases resembles an ovoid with $L_{ro} \approx 2.5D$ and $D_{ro} \approx 1.5D$. It should be noted that the exact body shape is insignificant as far as the wavelength of the waves is considered, but the wave amplitude {depends on the body shape}.

Fig. \ref{fig:lee_waves}(d)  contrasts the wave amplitude  between simulation and the theoretical estimate of  Eq. \ref{w_lee_waves}. The comparison is   along the dashed line with an arrow in Fig. \ref{fig:lee_waves}(a), which is at an angle of $45^{\circ}$ to the $x$-axis.
 The theory, when used with the Rankine ovoid approximation for the disk and the separated flow behind it, is able to accurately capture the amplitude variation specially moving away from the body for the 111 case which has linear stratification with constant $N$.


A similar analysis is performed for the lee waves of 614 after taking the Froude number in the analysis to be that of the linearly stratified region, i.e. $Fr = 6.13$ for Fig. \ref{fig:lee_waves}(e)  and $Fr = 3.74$ for Fig. \ref{fig:lee_waves}(f).  This procedure results in the correct prediction of the wavelength of the propagating lee waves. Theory  is able to capture the order of magnitude of the significantly reduced (relative to 111) wave amplitude but the  theoretical estimate of the wave amplitude is an under-prediction.
The disk moves through a local stratification corresponding to $\Fro = 1$ but the lee waves form and propagate in the linearly stratified regions where the Froude number is larger ($\Fro = 6.13$ and $\Fro = 3.74$).  Therefore, the true wave amplitude corresponds to $\Fro$ somewhat lower than that used  in the theory and, thus, larger than the theoretical estimate. Since the wavelength is comparable to the variability scale of $N(z)$, WKB theory cannot be used and we do not proceed further with linear analysis.

\subsection{Mean defect velocity}\label{section:ud}

\begin{figure}

\centerline{\includegraphics[width=\linewidth, keepaspectratio]{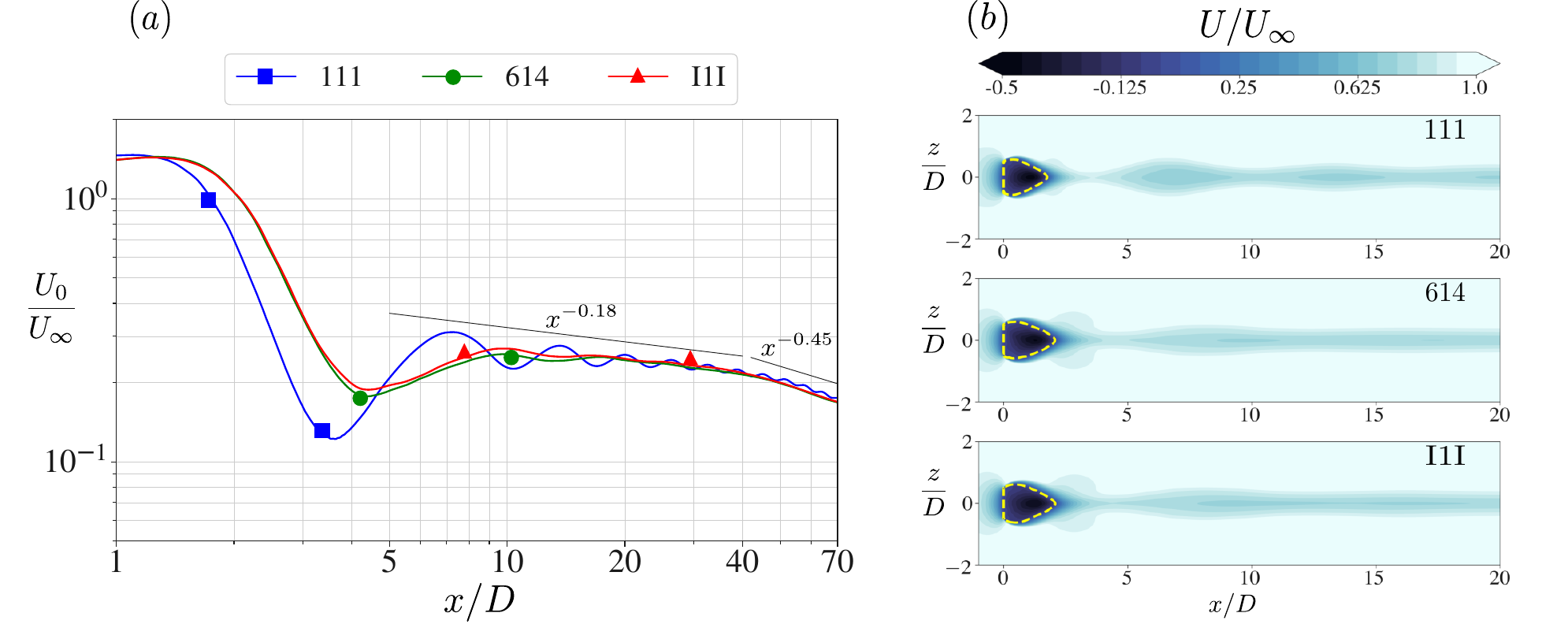}}
\caption{(a) Mean defect velocity at centerline. (b) Mean streamwise velocity contours on vertical plane passing through centerline. {Dashed yellow line represents the separation bubble ($U=0$).}}
\label{fig:ud}
\end{figure}

All cases show a strong effect of buoyancy on the wake since $\Fro = 1$ at the disk is common to them but there are differences among the cases as elaborated below. Fig. \ref{fig:ud}(a) shows the mean centerline defect velocity for the three cases plotted against the streamwise distance. The 111 case shows strong oscillatory modulation of the wake and its defect velocity, similar to the sphere wake~\citep{pal_direct_2017}.     In contrast to  111,  {the modulation in 614 and I1I, although present,  is  small}. The wavelength of modulation for 111 is $2 \pi D Fr$ as in the linear-theory  result (Eq. \ref{lee_waves}), and as noted in previous work. 

All cases show a similar decay law of $U_0 \propto x^{-0.18}$ in the NEQ regime, which was also observed by \cite{chongsiripinyo_decay_2020} for a disk at a higher Reynolds number of $5 \times 10^{4}$. The decay law transitions to $x^{-0.45}$ around $x/D = 50 (Nt = 50)$ {signaling} the beginning of the Q2D regime.  A notable difference in the defect velocity profiles occurs at  $x/D \sim 4$, where  the initial dip in the profile for 111 is larger in magnitude and occurs earlier than that of 614 and I1I, owing to the stronger lee wave in 111 as compared to 614 and I1I. The mean velocity contours on a vertical streamwise cut (Fig. \ref{fig:ud}b), show that the lee wave modulation in the wake for 111 is stronger and its  separation bubble is  smaller, which is consistent with the earlier dip in the profile of defect velocity (Fig. \ref{fig:ud}a).

\subsection{Turbulent kinetic energy}

\begin{figure}=

\centerline{\includegraphics[width=0.9\linewidth, keepaspectratio]{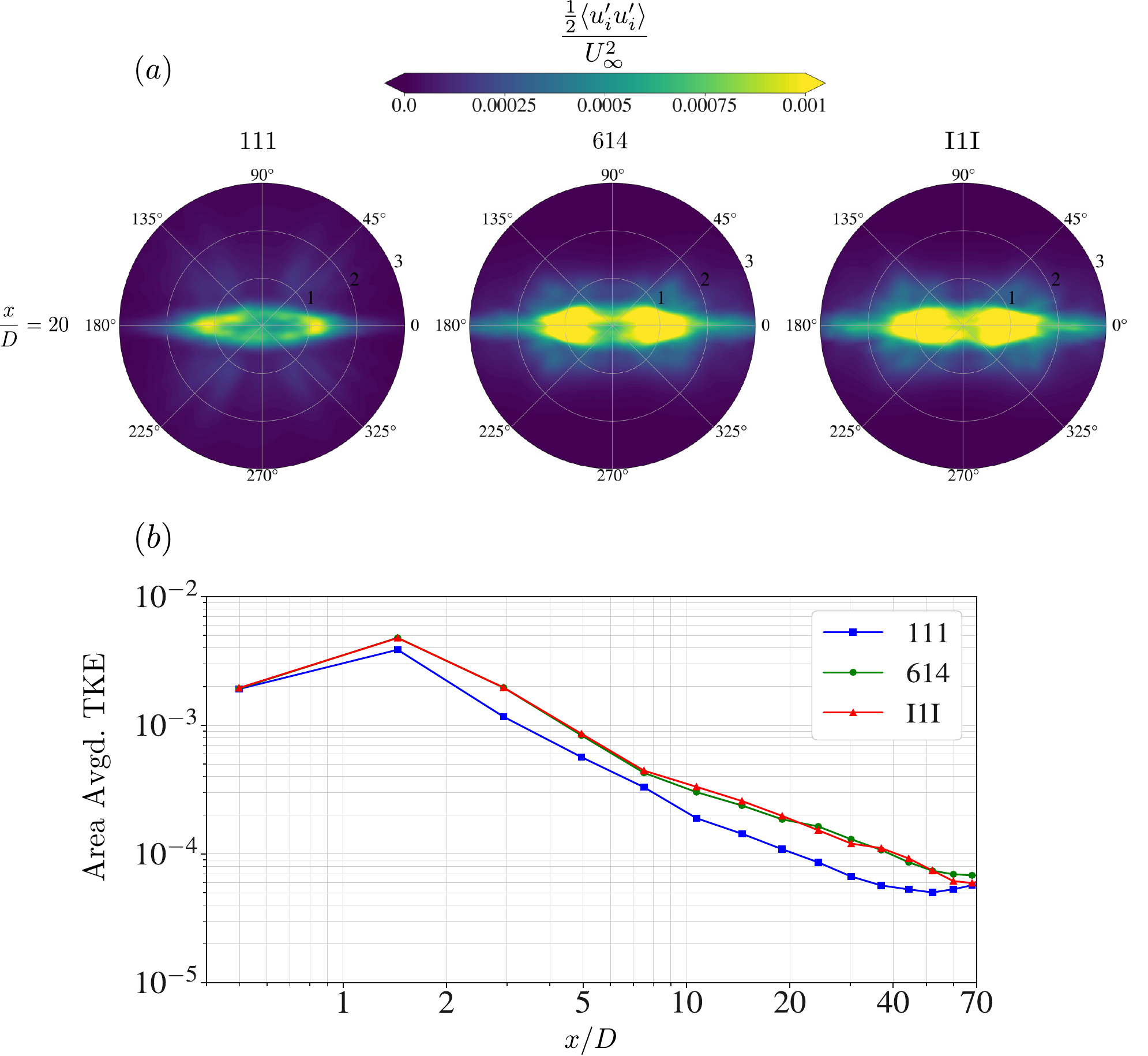}}
\caption{(a) Turbulent kinetic energy contours at $x/D = 20$. (b) Area averaged values of TKE as a function of streamwise distance.}
\label{fig:tke_cont}
\end{figure}

The nonlinearly stratified cases exhibit stronger levels of turbulent kinetic energy (TKE). This difference is illustrated by the contours of TKE at $x/D = 20$ in Fig. \ref{fig:tke_cont}(a), plotted for the three stratification profiles. The pycnocline cases have a higher TKE (by 50 - 100 \%) relative to the 111 case over the entire wake length as shown by the area-averaged values of TKE plotted in Fig. \ref{fig:tke_cont}(b). The area averaging at any streamwise location is done in the region covered by a circle of radius $3D$,
therefore containing most of the turbulent zone. The TKE is substantially larger in I1I and 614 relative to 111, especially in the NEQ regime.  For example,  I1I has almost twice the TKE of 111 at $x/D = 45$. {It is worth noting the slight increase in TKE for the case 111 at $x/D \approx 50$ which is where the decay law for the mean defect velocity changes (Section \ref{section:ud}), again pointing towards the transition to Q2D regime.}

 To explain the relatively high turbulence, we inspect  the terms in the TKE budget, especially the lateral production (the production by vertical Reynolds shear stress is small) and the wave flux. Although the hyperbolic tangent profile is close to a linear profile near the centerline, the weaker stratification regions above and below  are not as effective in suppressing near-wake turbulence stresses, which results in higher TKE production further downstream. 
 Also, the weaker far-field stratification in 614 and I1I relative to 111 {does} not allow the full  frequency range of  waves generated in the $Fr = 1$ central region to  escape into the far field. The reduction of wave energy flux  traps fluctuations in the wake and results in a TKE increase.
The two quantities, lateral TKE production and wave flux, are diagnosed in the following subsections.

\subsubsection{TKE production}

\begin{figure}
\centering
\centerline{\includegraphics[width=0.9\linewidth, keepaspectratio]{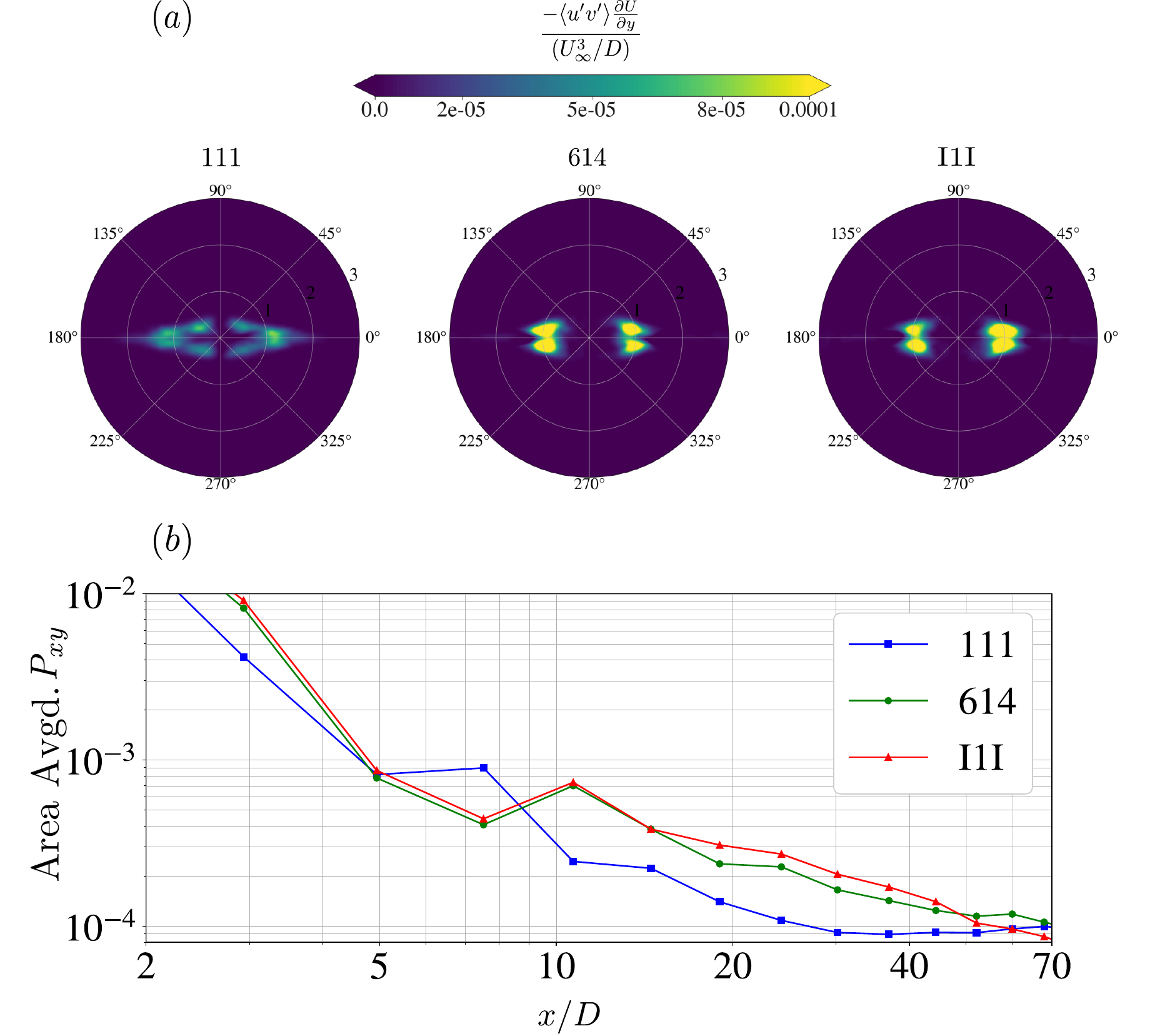}}
\caption{(a) Lateral production contours at $\frac{x}{D} = 20$. (b) Area-averaged values of lateral production as a function of streamwise distance. 
}
\label{fig:pxy}
\end{figure}

Contours of lateral production $P_{xy}$, which is the dominant component of TKE production for all three cases, are plotted in Fig. \ref{fig:pxy}(a) at $x/D = 20$. The increased $P_{xy}$ in 614 and I1I is partly a result of the higher lateral Reynolds shear stress $\langle-u'v'\rangle$. 

Area averaged-values of the lateral production over a circle of radius $3D$ are shown in Fig. \ref{fig:pxy}(b). The lateral production is consistently larger in the 614 and I1I cases over $10 < x/D < 50$ and is  as much as twice at some locations.

\subsubsection{Energy flux of wake generated unsteady waves}

The dispersion relation for internal gravity waves  in a stratified non-rotating medium, 
\begin{equation}
\Omega = N \cos {\Theta} \, , 
\label{eq:dispersion}
\end{equation}
where $\Theta$ is the angle of the phase line with the vertical,  limits the maximum allowable frequency to $N$.
\begin{figure}

\centerline{\includegraphics[width=0.9\linewidth, keepaspectratio]{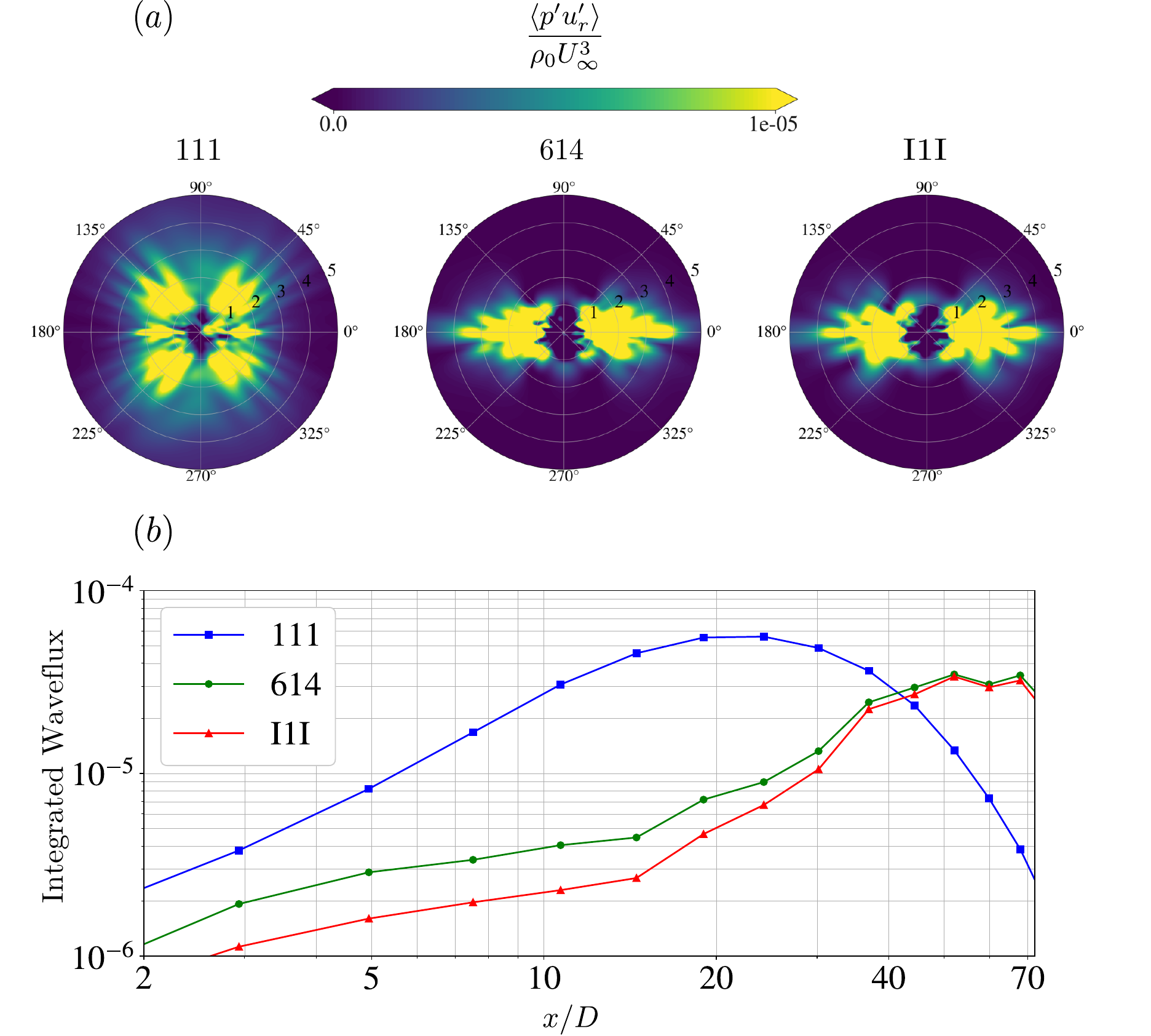}}
\caption{(a) Radial waveflux contours at $\frac{x}{D} = 20$. (b) Total waveflux integrated over rectangular perimeter as a function of streamwise distance.}
\label{fig:rwf}
\end{figure}
Since weaker stratification supports a narrower frequency band of  wave propagation,   614 and I1I  can be expected to allow a smaller  wave energy flux to radiate through the pycnocline layer as compared to 111. This is indeed the case as can be seen by the contours of the radial wave flux $\langle p' u_{r} ' \rangle$ in Fig. \ref{fig:rwf}(a), where  it is evident that the wave flux  is restricted within the pycnocline layer ($-2 < z/D < 2$) instead of being radiating away as in 111. Note that the mean of the radial velocity and pressure are subtracted when computing the flux so as to discard the component from the steady lee waves.

\begin{figure}

\centerline{\includegraphics[width=\linewidth, keepaspectratio]{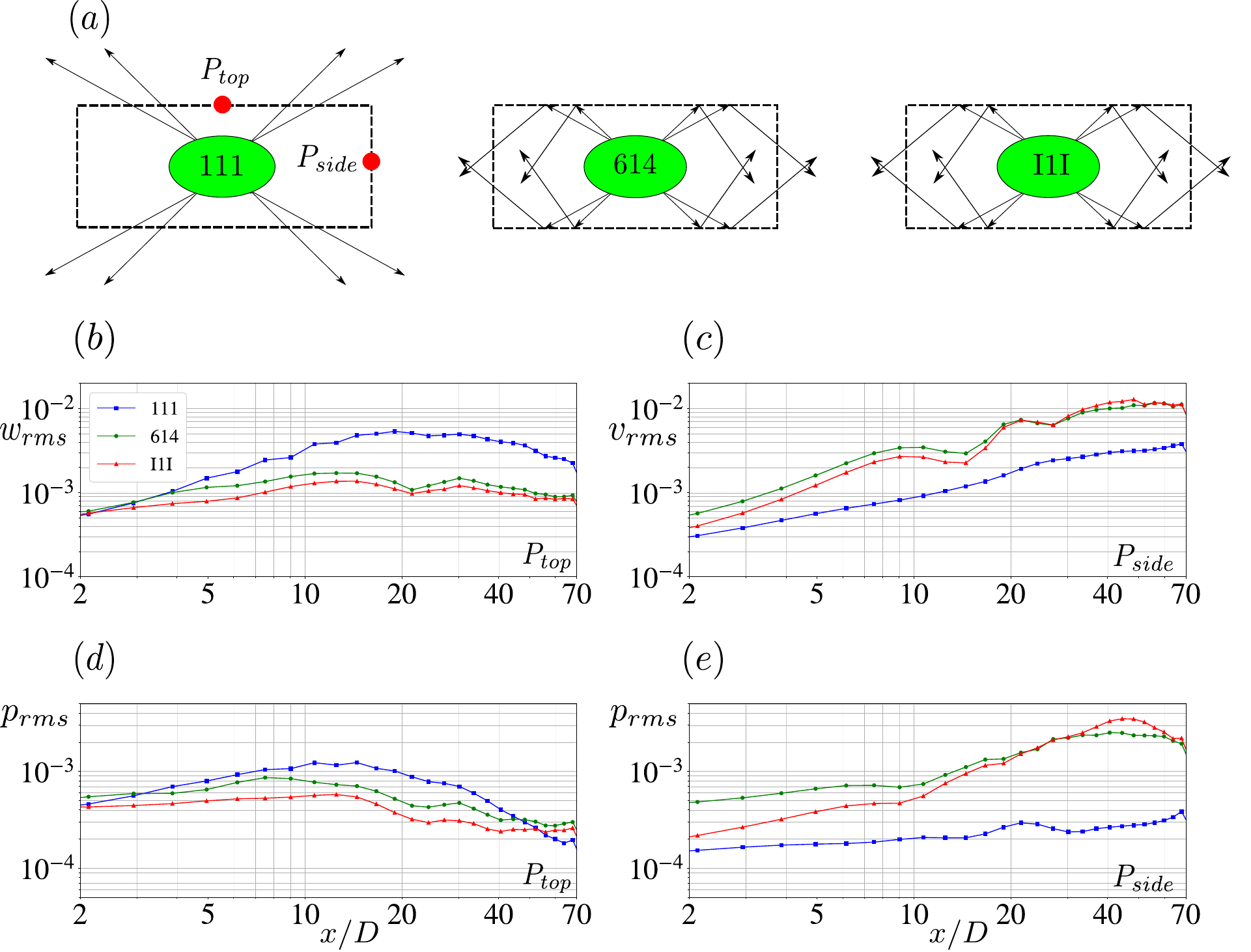}}
\caption{(a) Schematic showing trapping of internal waves for 614 and I1I compared to 111 (b) $w_{rms}$ at $P_{top}$ (c) $v_{rms}$ at $P_{side}$ (d) $p_{rms}$ at $P_{top}$ (e) $p_{rms}$ at $P_{side}$. 
}
\label{fig:pwv_rms}

\end{figure}

A quantitative comparison among the cases  is performed by computing the line integral of the internal wave flux over a rectangular perimeter. Specifically, the integral
\begin{equation}
\oint_{C} \frac{1}{\rho_{0}} \langle {\bf u}' p' \rangle . {\bf n}  \, dl 
\end{equation}
which is a function of $x/D$, is calculated using a rectangle  $C$  with  vertical sides at $y/D  = \pm 4$ and  top/bottom sides at $z/D = \pm 2.5$ ({thereby containing the pycnocline layer}). Fig.\ref{fig:rwf}(b) shows that  the flux of  wave energy  radiated away from the wake of 111 is much  larger than 614 and I1I  in the early part of the NEQ regime. However, at $x/D = 40$ in the late NEQ stage, the fluxes for 614 and I1I overtake the flux of 111.  Both of these trends can be explained by the schematic shown in Fig. \ref{fig:pwv_rms}(a). The wake generated internal waves that are trapped inside undergo complex wave-wave interaction after getting vertically restricted, resulting in sideward escape of the wave flux. The sideward escape can be verified by using an appropriate scaling argument for the vertical as well as lateral wave flux at the top and side boundaries of the rectangle, respectively.
 
 \begin{equation}
\langle p'w' \rangle \sim p_{rms} w_{rms},
\langle p'v' \rangle \sim p_{rms} v_{rms}
\end{equation}

 The values of $w_{rms}$ at the point $P_{top}$, $v_{rms}$ at the point $P_{side}$ and $p_{rms}$ at $P_{top}$ and $P_{side}$ are shown in Fig. \ref{fig:pwv_rms}(b-e), respectively. Comparison of Fig. \ref{fig:pwv_rms}(d) and \ref{fig:pwv_rms}(e) shows that,  although pressure fluctuations at the top boundary are highest  initially for 111 among all cases and boundaries,  the  pressure fluctuations for 614 and I1I increase with time at the side boundaries because of the wave trapping and the resulting sideward escape. The velocity fluctuations at the boundaries also show similar trends. The sideward escape of the wave energy takes some time to manifest which, in the simulation frame, 
  translates to streamwise distance, hence causing the slight delay in the increase of integrated wave flux for 614 and I1I.  Fig. \ref{fig:pwv_rms}(b-e) are consistent with the following result: most of the wave flux for 614 and I1I that escapes the wake core does so in the form of lateral wave flux $\langle p'v' \rangle$ unlike 111 where the vertical wave flux $\langle p'w' \rangle$ is the major contributor. 

\section{Centered vs. shifted nonlinear stratifications}\label{part2}

\subsection{Taylor-Goldstein Equation and Kelvin wake waves}\label{kelvin_waves}

In cases involving pycnoclines, there is a family of steady waves which resemble a Kelvin ship wave pattern, on horizontal planes inside the pycnocline layer. Intuitively, a  stratification profile with large central  value of buoyancy frequency ($\Fro \leq O(1)$) that changes rapidly in the vertical  (over $\delta/D \leq  O(1)$) to a small value at the edges can be thought of as a smoothed density jump and thus one can expect a pattern resembling that generated  on the air-water interface by a moving ship. Both centered and shifted stratification profiles show Kelvin wake wave patterns. However, as demonstrated here, the wake wave structure differs qualitatively from the  centered 614 wake when the stratification profile is shifted with respect to the disk center as in $614-$ and $614+$ .

The dispersion relation of air-water surface gravity waves leads to the waveforms of air-water Kelvin ship waves. To obtain an appropriate dispersion relation  for the nonlinear continuously stratified profiles,  the Taylor-Goldstein Equation is numerically  solved:

\begin{equation}
\frac{d^{2} \phi (z)}{d z^{2}} + k^{2} \Big(\frac{N^{2}(z)}{\Omega^{2}} - 1\Big)\phi (z) = 0.
\label{tge}
\end{equation} 
Here, $\phi(z)$ is the vertical displacement eigenfunction, and $k$ and $\Omega(k)$ are the wavenumber and angular frequency corresponding to the dispersion relation. The TGE is numerically solved as an eigenvalue problem for $\phi(z)$ by treating $\Omega$ as an eigenvalue and $k$ as a parameter, eg. by \cite{robey_thermocline_1997}. The boundary conditions used are 

\begin{equation}
\phi(z=-L_{r}) = \phi(z=L_{r}) = 0.
\end{equation}



Also, based on Sturm-Liouville theory, the eigenfunctions are normalized using:
\begin{equation}
\int_{z=-L_{r}}^{z=L_{r}} \phi_{m} N^{2}(z) \phi_{n} \,dz = \delta_{mn}.
\end{equation}

\begin{figure}

\centerline{\includegraphics[width=\linewidth, keepaspectratio]{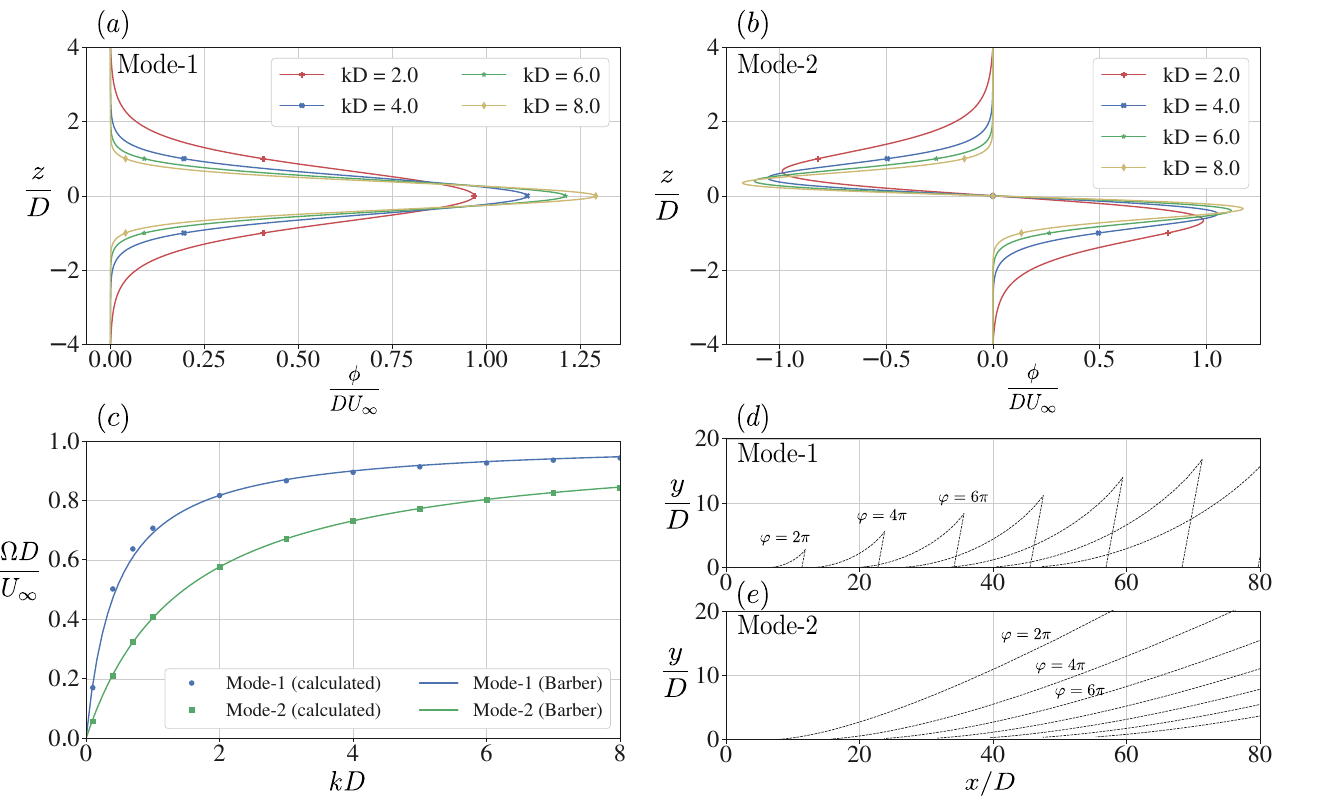}}
\caption{(a) Mode-1 and (b) Mode-2 eigenfunctions for different values of wavenumber in Eq. \ref{tge}. (c) Numerically calculated dispersion relation compared with approximation given by \cite{barber_dispersion_1993}.(d) Mode-1 and (e) Mode-2 waveforms as calculated using Eq. \ref{wave_pattern}.}

\label{fig:eigfns}
\end{figure}

Fig. \ref{fig:eigfns}(a) and \ref{fig:eigfns}(b) respectively show the mode 1 and mode 2 eigenfunctions for different values of $k$. The numerically calculated dispersion relation is also compared in Fig. \ref{fig:eigfns}(c) to the approximation given by \cite{barber_dispersion_1993}:
\begin{equation}
\Omega(k) = \frac{c_{p_{0}}k}{1 + \frac{c_{p_{0}}k}{N_{max}}}.
\label{dr_approx}
\end{equation}
Here, $c_{p_{0}}$ is the limiting long-wave phase speed of a given mode at $k=0$, which is obtained by extrapolation, i.e., $c_{p_{0}}$ is the slope at the origin of the mode-specific curve in Fig. \ref{fig:eigfns}(c). For the chosen profiles, $c_{p_{0}}/U_{\infty}$ for mode 1 and mode  2 are $2.23$ and $ 0.68$, respectively.
Since Eq. \ref{dr_approx} provides a very good approximation to the dispersion relation,  it will be used for the remaining analysis of this section.

Using the dispersion relation, phase and group velocities associated with each mode can be calculated using $c_{p} = \Omega / k$ and $c_{g} = d\Omega / dk$, and then substituted in the expressions given by \cite{keller_internal_1970} to obtain the modal wave patterns corresponding to each eigenfunction:
\begin{equation}
x = \frac{\varphi U_{\infty} \Big(1 - \frac{c_{p} c_{g}}{U_{\infty}^{2}}\Big)}{k(c_{p}-c_{g})}, y = \frac{\varphi c_{g} \Big(1 - \frac{c_{p}^{2}}{U_{\infty}^{2}}\Big)^{\frac{1}{2}}}{k(c_{p}-c_{g})} ,
\label{wave_pattern}
\end{equation}
where $(x,y)$ corresponds to a locus of points on the horizontal plane parametrically given in terms of $k$ for a constant value of phase, $\varphi$. Fig. \ref{fig:eigfns}(d) and \ref{fig:eigfns}(e) respectively show mode 1 and mode 2 wave patterns, respectively,  plotted using Eq. \ref{wave_pattern}. 

\begin{figure}

\centerline{\includegraphics[width=0.9\linewidth, keepaspectratio]{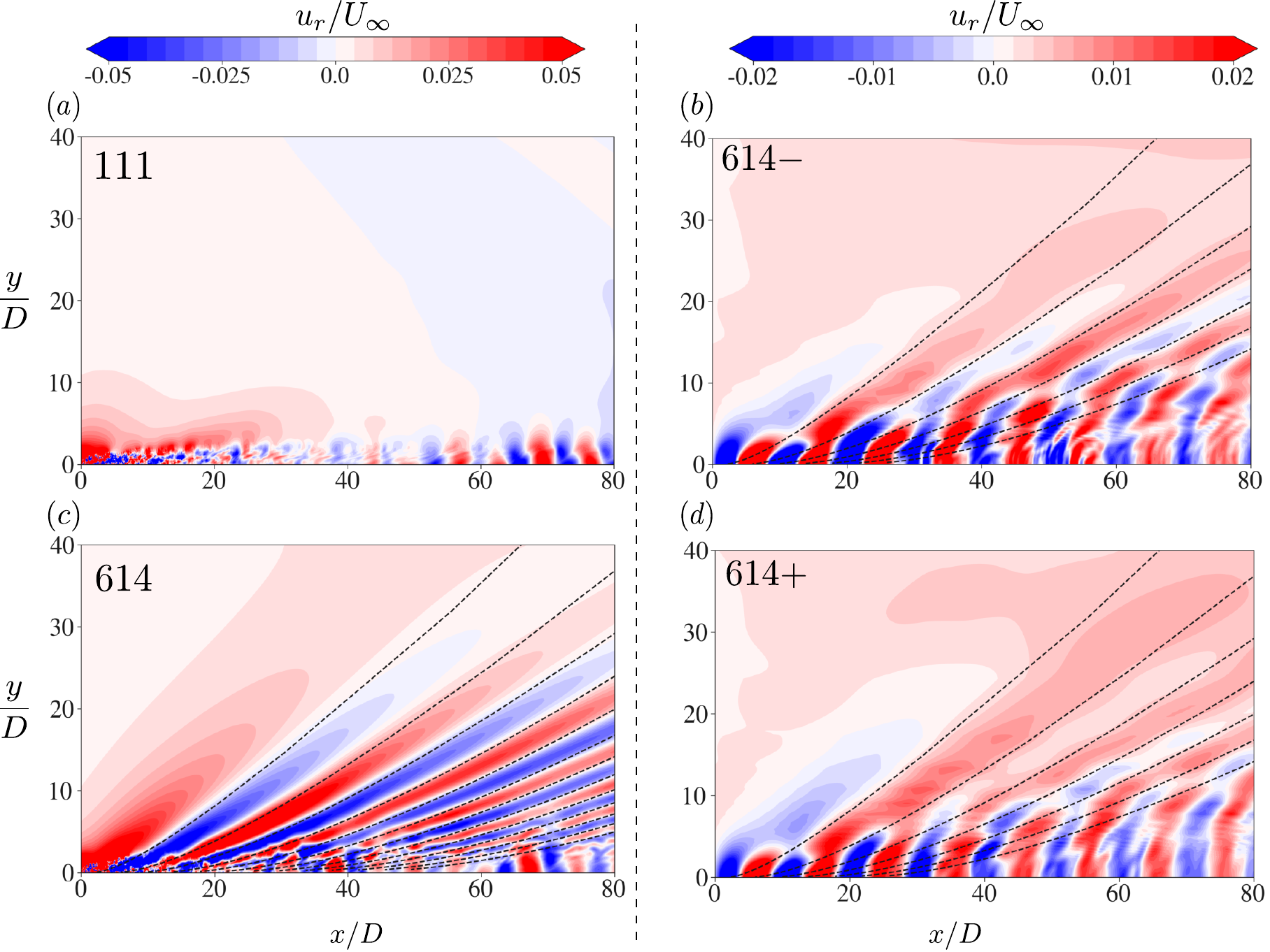}}
\caption{Instantaneous contours of radial velocity on a half-pycnocline-center plane:  (a) 111, (b) $614-$, (c) 614, and (d) $614+$. Dashed lines correspond to mode 2 waves which were shown in  Fig. \ref{fig:eigfns}(e). 
}
\label{fig:kelvin_waves}
\end{figure}

Fig. \ref{fig:kelvin_waves} shows the instantaneous radial velocity contours for 111, 614, $614-$ and $614+$ plotted on a half-horizontal plane passing through the respective pycnocline center. (For 111, the plot is on $\theta = 0^{\circ}$ plane). The waves in 614 have constant-phase lines which resemble mode 2 (shown earlier in Fig. \ref{fig:eigfns}e) but not mode 1.  In contrast, the phase lines for $614-$ and $614+$ are more complex.  In the region $0 < z < 10$ they resemble the mode 1 pattern of Fig. \ref{fig:eigfns}(d)  but, for $z > 10$, they resemble the mode 2 pattern. The manifestation of any mode by a disturbance in the pycnocline layer depends on the location of the disturbance with respect to the pycnocline layer. In 614, the disk center travels along the center of the pycnocline layer to displace fluid symmetrically in the upward and downward direction, corresponding to the antisymmetric mode 2 eigenfunction for displacement and therefore generates mode 2 waves. Any vertical offset from the pycnocline center modifies the antisymmetric pattern of the displacement so as to also involve some contribution from the mode 1 waveform. The presence of a mode 2 wave pattern in the central region of the lateral plane and a mode 1 pattern otherwise  is in agreement with the visualizations in the experiment of \cite{robey_thermocline_1997}. For profile $614+$, the contours look similar to that of $614-$ but with slightly lower intensity because of the disk being further away from the center of the pycnocline. (Note that the wavepattern for I1I is the same as that of 614 because they have the same hyperbolic tangent function modeling the nonlinearity.)

\subsection{Distinction between lee waves and Kelvin wake waves}

\begin{figure}

\centerline{\includegraphics[width=\linewidth, keepaspectratio]{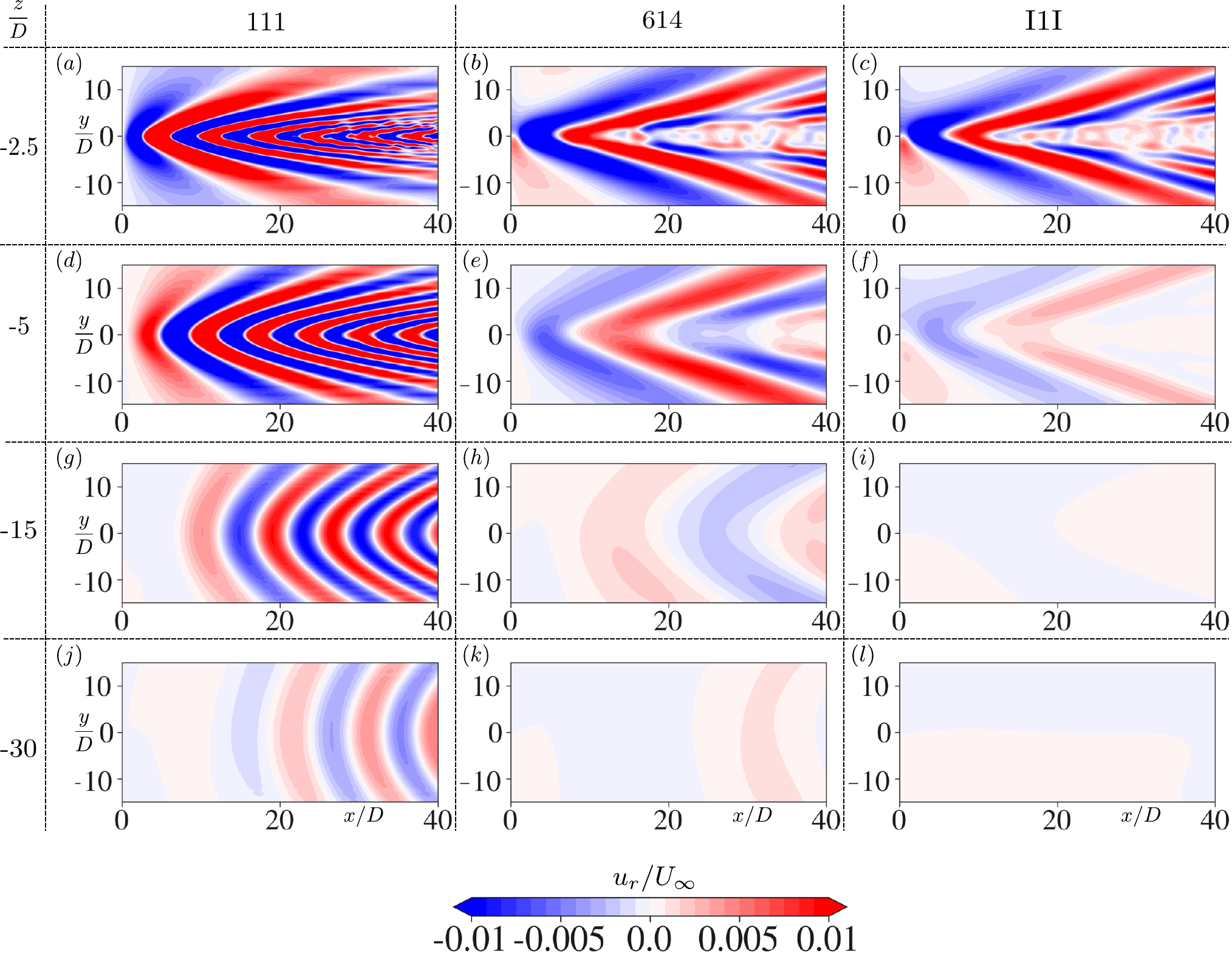}}
\caption{Instantaneous contours of radial velocity for the centered cases on vertically offset horizontal planes for  111 (left column), 614 (middle) and I1I (right). The vertical offset  of each plane increases from the top to bottom row. 
}
\label{fig:bothwaves}
\end{figure}

Kelvin wake waves are steady in the frame of reference of the moving body like the lee waves but constitute a distinct family of waves and appear only when the stratification is nonuniform.
Fig. \ref{fig:bothwaves} shows instantaneous contours of radial velocity on horizontal planes for the centered profiles 111, 614, and I1I at different vertical locations. At $z/D=-2.5$ (Fig. \ref{fig:bothwaves} (a-c)), the contours have superficial resemblance but they represent fundamentally different waves. For 111, since there is no nonlinear gradient, Kelvin wake waves are not formed on the horizontal plane passing through the domain centerline at $z/D=0$, (e.g. the previously shown Fig. \ref{fig:kelvin_waves}a), and so Fig. \ref{fig:bothwaves}(a)  has the horizontal imprint of only lee waves. 
 However, for 614 and I1I, the contours at $z/D=-2.5$ in Fig. \ref{fig:bothwaves}(b,c) are mostly the imprints of the Kelvin wake waves in Fig. \ref{fig:kelvin_waves}(c). For 614, Kelvin wake waves disappear farther away from the region of nonlinear stratification and the contours transition to those of lee waves corresponding to $\Fro=3.74$ (Fig. \ref{fig:bothwaves}(e,h,k)). For I1I, once the Kelvin wake waves disappear, lee waves are not observed since the profile is unstratified outside the pycnocline region (Fig. \ref{fig:bothwaves}(f,i,l)). 
Turning back to case 111, we see the lee waves weakening as $z/D$ is increased (Fig. \ref{fig:bothwaves}(d,g,j)) but their pattern is still clear at $z/D = -30$.


\subsection{Wake generated internal gravity waves}

\begin{figure}

\centerline{\includegraphics[width=0.8\linewidth, keepaspectratio]{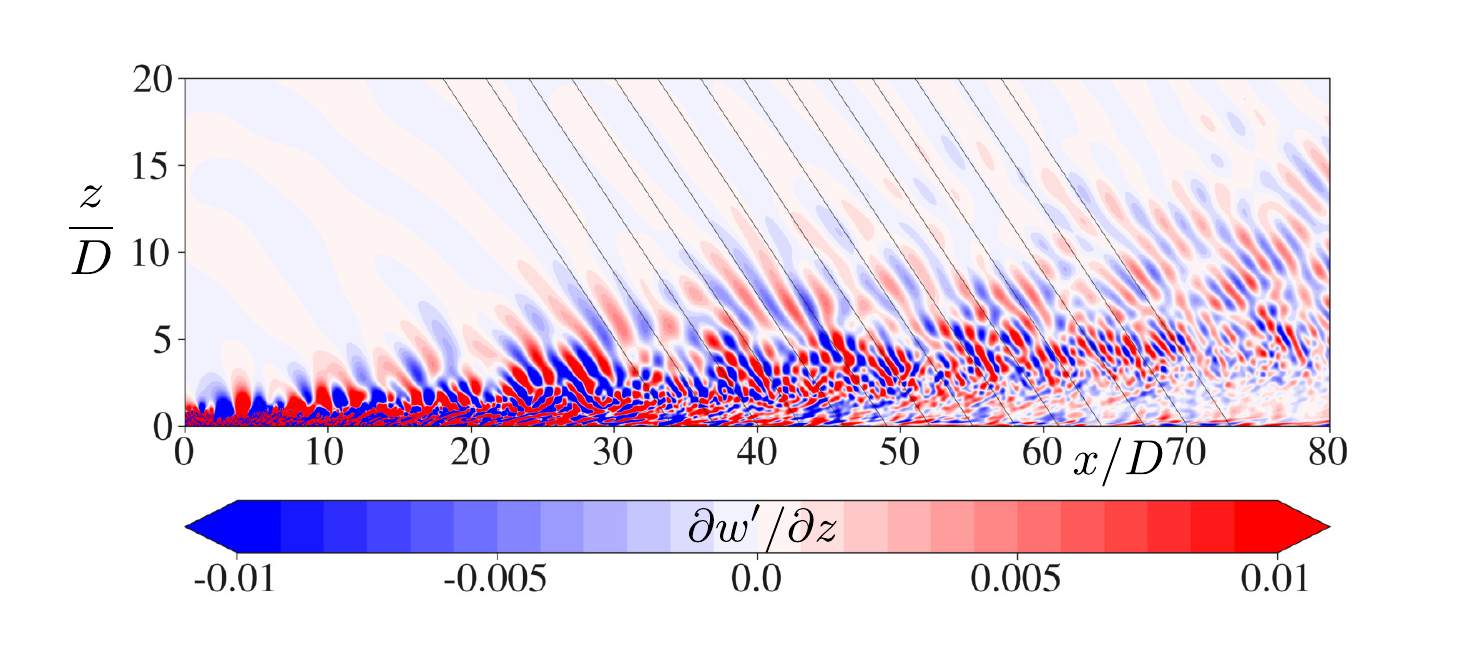}}
\caption{$\partial w' / \partial z$ contours showing wake generated internal gravity waves on the $\theta = 90^{\circ}$ plane for 111. Black lines are inclined at an angle of $39^{\circ}$ with the vertical.}
\label{fig:111waves}

\end{figure}




The  internal gravity waves generated by the turbulent wake are unsteady. These waves are visualized in Fig. \ref{fig:111waves} for case 111 by contours of $\partial w / \partial z$ on the $\theta = 90^{\circ}$ vertical plane. The phase lines of the radiated waves in the far field are seen to cluster around a  a characteristic inclination angle which suggests a narrow frequency band in the far field according to the dispersion relationship, Eq. \ref{eq:dispersion}, for the intrinsic  (in a frame where the background has zero velocity) wave frequency.
Fig. \ref{fig:111waves} also shows solid black lines plotted at an angle of $39^{\circ}$ from the vertical, which also seems to be the angle for the waves ($39^{\circ} \pm 2^{\circ}$). This gives $\Omega / N = 0.78 \pm 0.03$.  

\begin{figure}

\centerline{\includegraphics[width=0.8\linewidth, keepaspectratio]{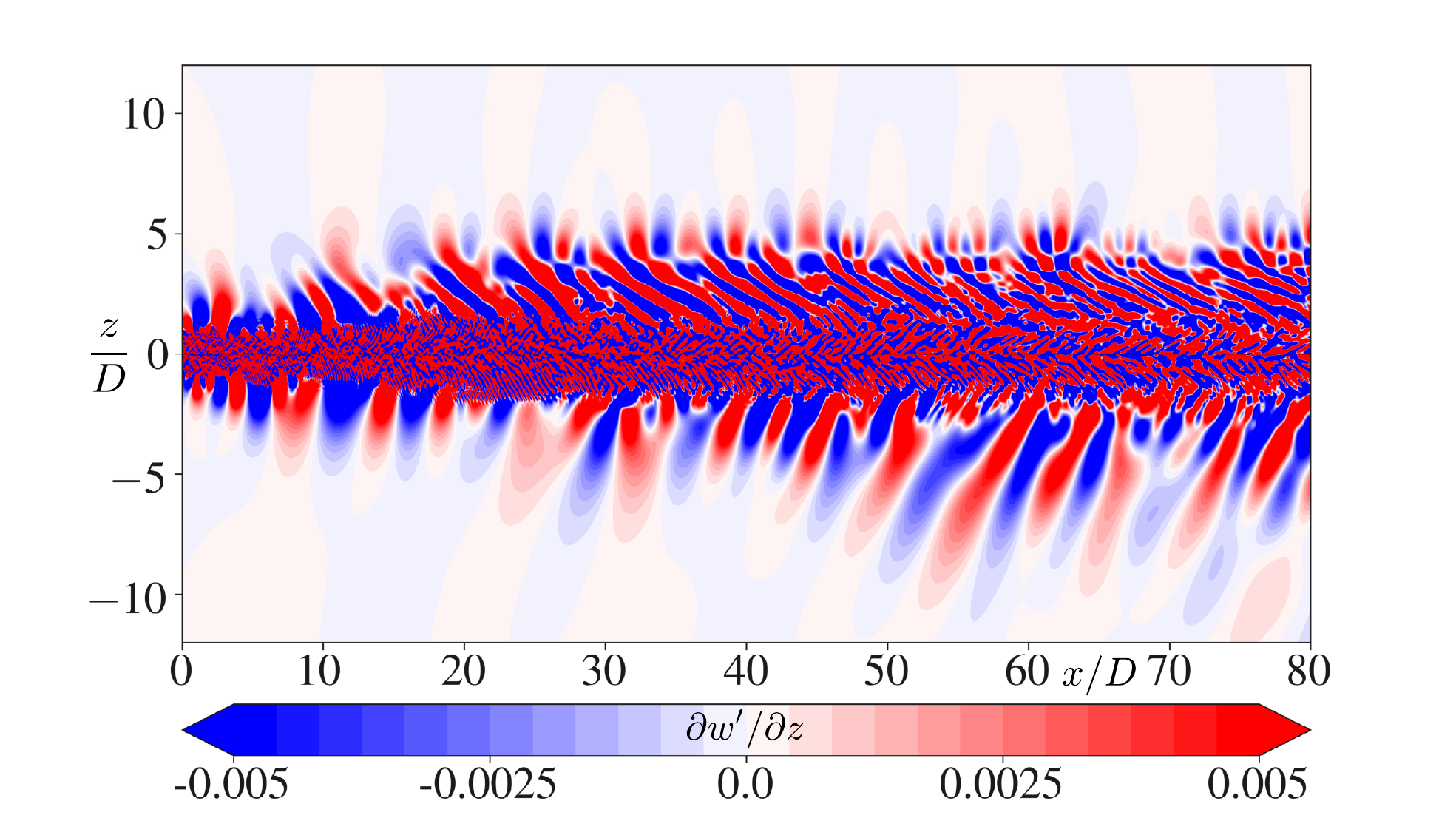}}
\caption{$\partial w' / \partial z$ contours showing wake generated internal gravity waves on the $\theta = 90^{\circ}$ and $180^{\circ}$plane for $614+$.}

\label{fig:s2waves}

\end{figure}



For profile $614+$, the $\partial w' / \partial z$ contours on the entire vertical plane ($\theta = 90^{\circ}$ and $270^{\circ}$) are plotted in Fig. \ref{fig:s2waves}. 
Phase lines with a dominant inclination angle are seen in the negative $z/D$ region where the waves   with downward group velocity travel in an entirely linear stratification ($\Fro = 3.74$). The inclination angles in this region of $z/D < 0$  cluster around $\Theta = 44^{\circ}$ which gives $\Omega / N \approx  0.72$ (Note that $N$ here corresponds to the local $\Fro = 3.74$). The upward moving  waves generated by the disk enter a region of nonlinear stratification where there is significant small-scale variability. As the waves exit the pycnocline layer from the top, they have near-zero inclination with respect to the vertical, i.e. $\Theta \approx  0^{\circ}$, which implies that only the high-frequency near-$N$ ($N$ here again corresponds to the local $\Fro = 6.13$) part of the wake generated waves escapes into the upper region of weak stratification.  Thus, a significant portion of the wake generated waves is trapped within the pycnocline leading to the small-scale variability in that region.

\subsection{Wake characteristics}\label{wake}

\begin{figure}

\centerline{\includegraphics[width=0.7\linewidth, keepaspectratio]{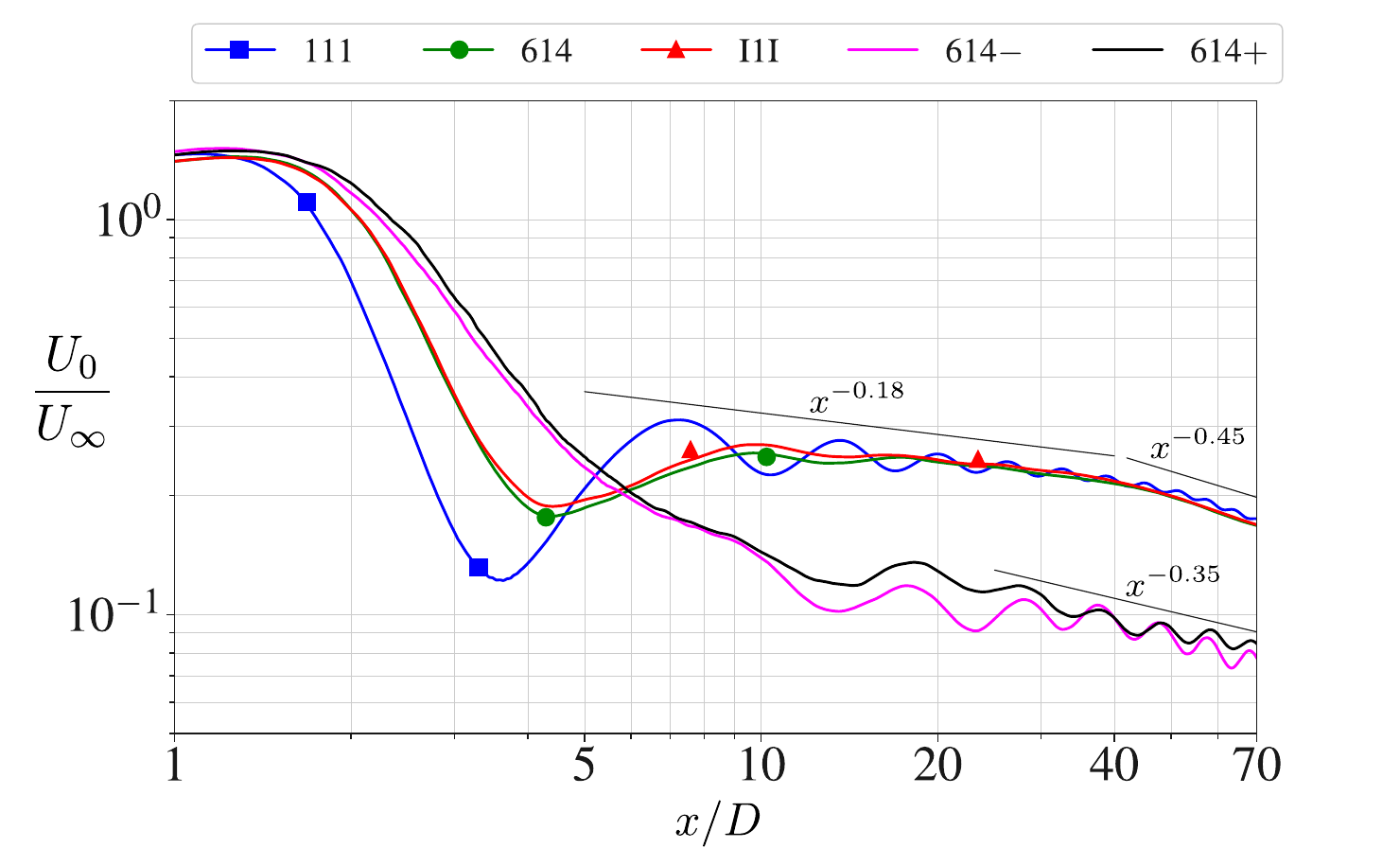}}
\caption{Evolution of centerline mean defect velocity for the five simulated cases. }
\label{fig:meanu_all_5}

\end{figure}

Wake characteristics, namely, the mean defect velocity and turbulent kinetic energy, show qualitatively different behavior   for the shifted pycnocline cases $614-$ and $614+$ relative to the centered 614 profile. Fig. \ref{fig:meanu_all_5} compares the streamwise  evolution of centerline mean defect velocity ($U_0/U_\infty$) among  all  five simulated cases. Recall that for $614+$, the disk is entirely in the $\Fro = 3.74$ region with its upper edge in the pycnocline and, for $614-$, half of the disk is in the $\Fro = 6.13$ region and half in the pycnocline. Since, for both $614-$ and $614+$,  the disk wake  evolves in relatively weaker stratification relative to the $Fr =1$ stratification seen by the disk in 614, there is a faster rate of decay of $U_0$ {($\propto x^{-0.35}$)} in these cases (Fig. \ref{fig:meanu_all_5}). $614-$ has a weaker effective stratification relative to $614+$ and, therefore, has a somewhat lower $U_0$. 

\begin{figure}

\centerline{\includegraphics[width=0.9\linewidth, keepaspectratio]{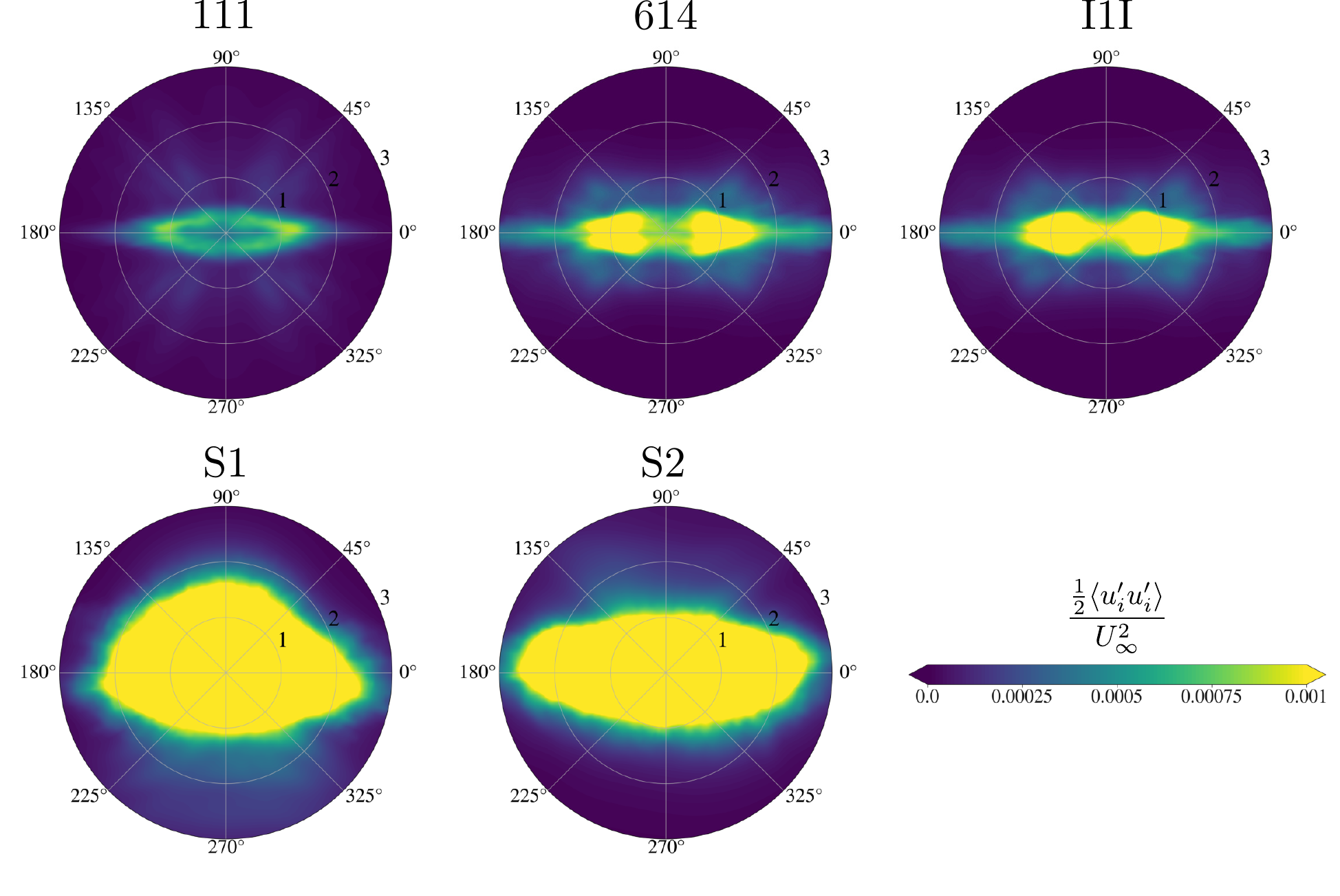}}
\caption{Turbulent kinetic energy contours at $x/D = 20$ for all five cases.}
\label{fig:tke_cont_all_5}

\end{figure}

The TKE contours plotted in Fig. \ref{fig:tke_cont_all_5} also show that the buoyancy effect is weaker for the shifted profiles. $614-$ and $614+$ show far higher TKE than the other three cases that have $\Fro = 1$ at the disk center. Since the bottom half of the disk of $614-$ is inside the pycnocline layer, the wake core is not symmetric about the horizontal axis and appears to be somewhat compressed from the bottom where the effective stratification is higher. The TKE profile at $x/D = 20$ is more symmetric for $614+$. 

\section{Summary and Conclusions} \label{conclusions}

We present results from a body-inclusive LES of turbulent wakes in a  background stratification which is nonlinear instead of the linear stratification that is typically considered. Four density profiles with a hyperbolic tangent pycnocline are selected along with a standard constant linear stratification profile. The nonlinear profiles are 614 (weak upper stratification with $Fr \approx 6$ and weak lower stratification with $Fr \approx 4$ bounding the pycnocline), I1I (no stratification surrounding pycnocline), $614-$ (614 vertically shifted by $-2.5D$), and $614+$ (614 vertically shifted by $3D$). A disk of diameter $D$ is chosen as the canonical bluff body, the relative velocity is $U_{\infty}$ and  the  Reynolds number is $5000$. The Froude number $Fr (z) = U_{\infty} /N(z) D$, which is based on local value of buoyancy frequency $N(z)$, varies and takes a   minimum value of  $1$ for the cases with nonuniform $N(z)$ and is set to 1 in the benchmark constant-$N$ case. The inverse of $Fr(z)$ can be viewed as the local buoyancy frequency, nondimensionalized with the flow scales, so that nondimensional $N_{max}$ is the same among all cases. The pycnocline thickness normalized by $D$ is also held constant. 


Our main conclusion is that  the non-constant $N$  profile studied here substantially alters both the turbulent wake and the internal wave field.    In the first part of the study, results with the nonlinear profiles 614 and I1I are compared with the linear profile 111.  The mean defect velocity ($U_0/U_\infty$) 
in  the near wake ($x/D < 8$) 
is quite different for 111 which shows an initial oscillation with  stronger amplitude than 614 and I1I.  This  initial oscillation  of the defect velocity is the oscillatory modulation of the wake,  which  has been shown to be an imprint of the steady lee wave field in linear stratification~\citep{pal_direct_2017}.  The I1I case, which is unstratified outside the pycnocline,  does not support a steady lee wave in the far field. Nevertheless, its near wake exhibits  oscillatory modulation (wavelength is slightly larger than that for 111) because of an evanescent steady wave.
In the   NEQ regime, which  follows the initial oscillation, $U_0/U_\infty$ in all three cases was found to decay as $x^{-0.18}$. \cite{chongsiripinyo_decay_2020} found  the same power-law exponent of $-0.18$ in  their disk wake, which had  an order of magnitude larger $\Rey = 5\times 10^4$.
 The decay rate transitions to $x^{-0.45}$ around $x/D = 45$, suggesting the end of the NEQ regime and beginning of the  regime with Q2D power law.  The Q2D regime was not found by \cite{chongsiripinyo_decay_2020} at their lowest Froude number, $\Fro = 2$. 
 
Wakes formed within a pycnocline layer are found here to be more turbulent than  in linear stratification given the same  value of $N(z)$ at the disk center. The TKE for cases 614 and I1I exceeds that for 111 even though  $N(z)$ is  very similar in the wake core among the three cases. The difference is specially large in the NEQ regime where I1I has about twice the TKE of 111. The reasons are as follows. First, the most dominant production term, $P_{xy}$, is higher for 614 and I1I. The turbulent momentum flux  is sensitive to the weak stratification, even though the weakening occurs away from and not in the wake core and, therefore,  the suppression of the turbulent flux by stabilizing buoyancy is weaker  in 614 and I1I.   Second, the internal wave energy flux  out of the wake, $\langle p'u_{r} ' \rangle$ is smaller for 614 and I1I because there is a range of internal waves that is generated but  cannot  propagate out vertically  since their frequency exceeds the smaller $N$ outside the pycnocline. This wave energy is trapped inside the pycnocline layer. Both of these effects dominate at the beginning of the NEQ regime,  resulting in a more turbulent wake. It is worth noting that,  a side lobe of wave radiation forms in 614 and I1I (center and left panels of Fig. \ref{fig:rwf}a), but the  overall integrated wave flux in the NEQ regime is still smaller than 111.

Based on the  theoretical asymptotic analysis for  steady lee waves by \cite{voisin_sphere_2007}, the potential flow solution for an oval was used by  \cite{ortiz_spheroid_2019} to develop an expression for the lee wave field of  a spheroid in linear stratification.  We find that the disk, along with its separation bubble, can be  approximated as a Rankine ovoid 
and have 
used the ovoid shape for the potential flow solution instead of an oval. There is excellent agreement between the 111  simulation and the analytical expression with regards to both wave length and wave amplitude. The 614 simulation has a lee wave field whose wavelength is well predicted by  constant-$N$  linear theory upon using $\Fro = 6.13$ above and   $\Fro =3.74$  below the pycnocline to approximate the steady lee wave in those regions.  However, the simulation amplitude for 614 is  higher than that given by this approximation  since the  wave-generation region, which includes  the pycnocline, has an effective $N$ that is larger then the constant far-field $N$ and, therefore, a stronger wave field. 

In the second part of the study, the influence of shifting the pycnocline layer relative to the disk center is found to be strong.  In 614-, half of the disk lies in the upper region of weak ($\Fro = 6.13$) linear stratification and, in 614+, the disk lies entirely in the lower linearly stratified region ($\Fro =3.74$) but still very close to the nonuniform-$N$ region. Overall, the relative shift between pycnocline and disk weakens the buoyancy effect on the mean wake, since the wake feels a weaker effective stratification. Thus,  different from the 614 centered case, the defect velocity in the shifted cases do not show an oscillation  in $x/D < 8$ and also decays more quickly than 614 in the far wake. The wakes are also more turbulent  in $614-$ and $614+$ as shown by the TKE. The asymmetric placement (with respect to the profile) of the disk in $614-$ is also clearly manifested in its TKE contour where the lower half, which is inside the pycnocline layer is vertically compressed   relative to the upper half.

With regards to the wave field,  in addition to  the usual steady far-field lee waves and unsteady  wake generated waves, the pycnocline supports  a third family of waves, the steady {Kelvin wake waves}.
The {Kelvin wake waves} can be analytically described by  solving the Taylor-Goldstein equation as an eigenvalue problem and they take a fundamentally different form than the  lee waves as visualized by Fig. \ref{fig:bothwaves}. Furthermore, shifting the disk alters the modal wave form.  The dominant waveform in $614$ corresponds to the mode-2 eigenfunction   while  $614-$ and $614+$ exhibit a mix of two waveforms, corresponding to mode-1 and mode-2 eigenfunction. 
 In 614, the disk  moves right at the center of the pycnocline layer leading to the  symmetry property of the mode-2 wave. Any vertical offset from the center leads to the appearance of mode-1 waves. These modal waveforms are well corroborated by the experiments of \cite{robey_thermocline_1997} as well as \cite{nicolaou_internal_1995}. 

Phase lines of unsteady  internal gravity waves in the linearly stratified far field are found to cluster around a characteristic inclination angle which takes the value of   $\theta \approx 39^\circ$ in 111. 
This result of phase-angle clustering is  consistent with several previous studies of wave radiation from a turbulent flow \citep{sutherland_internal_1998, dohan_internal_2003, taylor_internal_2007, pham_dynamics_2009}  but it is also worth noting that as  $\Rey$ increases, \cite{abdilghanie_internal_2013} find   a broader range of wave phase angles.    In case $614+$, the wake generated waves that propagate downward do so in a linear stratification and cluster around $\theta \approx 44^\circ$. The upward propagating waves have to go through the pycnocline layer and only the near-$N$ waves are able to propagate   in the weak linear stratification above the pycnocline. Some of the wave energy that reflects off the pycnocline boundaries is trapped and leads to small-scale variability while a fraction escapes  through the sides of the wake. 


We have explored a limited portion of  the parameter space applicable to turbulent wakes in nonlinear background stratification. Depending on the application and the environmental setting, there can be wide variability in the  nondimensional pycnocline thickness ($\delta/D$), the shape of the nonuniform stratification, and the variability of $N(z)$.  Body characteristics such as its nominal Froude number, Reynolds number and its shape can vary. Future examination of this wide parameter space in laboratory experiments as well as simulations (body-inclusive hybrid spatial or temporal as well as the cheaper body-exclusive temporal with a good choice for initial conditions) will be useful.


\backsection[Funding]{The authors gratefully acknowledge the support of Office of Naval Research Grant N00014-20-1-2253.}

\backsection[Declaration of interests] {The authors report no conflict of interest.}

\bibliographystyle{jfm}
\bibliography{disk_pycnocline_re5k}
\end{document}